\documentclass[a4paper]{cas-sc}
\usepackage{doi}
\usepackage{lineno}
\usepackage{blindtext}
\usepackage{hyperref}
\usepackage[numbers,sort&compress]{natbib}
\usepackage{amsmath,amssymb}
\usepackage{amsthm}
\usepackage{mathtools}
\usepackage{paralist}
\usepackage{subcaption}
\usepackage{float}
\usepackage{float}
\usepackage[subnum]{cases}
\usepackage{xcolor}
\usepackage{graphicx}
\usepackage{algorithm}
\usepackage{algorithmicx}
\usepackage{algcompatible}
\DeclareMathAlphabet{\mathcal}{OMS}{cmsy}{m}{n}   
\makeatletter
\newcommand*{\rom}[1]{\expandafter\@slowromancap\romannumeral #1@}
\makeatother

\makeatletter

\makeatletter
\def\thickhline{%
  \noalign{\ifnum0=`}\fi\hrule \@height \thickarrayrulewidth \futurelet
   \reserved@a\@xthickhline}
\def\@xthickhline{\ifx\reserved@a\thickhline
               \vskip\doublerulesep
               \vskip-\thickarrayrulewidth
             \fi
      \ifnum0=`{\fi}}
\makeatother

\newlength{\thickarrayrulewidth}
\xspaceaddexceptions{]}
\def\tsc#1{\csdef{#1}{\textsc{\lowercase{#1}}\xspace}}
\tsc{WGM}
\tsc{QE}
\tsc{EP}
\tsc{PMS}
\tsc{BEC}
\tsc{DE}

\begin{document}\sloppy
\makeatletter
\renewcommand\section{\@startsection{section}{1}{\z@}%
    {15pt \@plus 3\p@ \@minus 3\p@}%
    {4\p@}%
    {
     \sectionfont\raggedright\hst[13pt]}}
\renewcommand\subsection{\@startsection{subsection}{2}{\z@}%
    {10pt \@plus 3\p@ \@minus 2\p@}%
    {.1\p@}%
    {
     \ssectionfont\raggedright }}
\makeatother

\let\WriteBookmarks\relax
\def\floatpagepagefraction{1}
\def\textpagefraction{.001}
\shorttitle{\emph{S. He \textit{et al.}}}
\shortauthors{S. He \textit{et al.}}

\title [mode = title]{Incorporating lane-change prediction into energy-efficient speed control of connected autonomous vehicles at intersections}


\author[1]{Maziar Zamanpour}
\ead{zaman050@umn.edu}
\author[2]{Suiyi He}
\ead{he000231@umn.edu}
\author[1]{Michael W. Levin}
\ead{mlevin@umn.edu}
\author[2]{Zongxuan Sun}
\ead{zsun@umn.edu}

\address[1]{Department of Civil, Environmental, and Geo-Engineering, University of Minnesota, Minneapolis, MN 55455, USA}
\address[2]{Department of Mechanical Engineering, University of Minnesota, Minneapolis, MN 55455, USA}

\begin{abstract}[S U M M A R Y]
Connected and autonomous vehicles (CAVs) possess the capability of perception and information broadcasting with other CAVs and connected intersections. Additionally, they exhibit computational abilities and can be controlled strategically, offering energy benefits. One potential control strategy is real-time speed control, which adjusts the vehicle speed by taking advantage of broadcasted traffic information, such as signal timings. However, the optimal control is likely to increase the gap in front of the controlled CAV, which induces lane changing by other drivers. This study proposes a modified traffic flow model that aims to predict lane-changing occurrences and assess the impact of lane changes on future traffic states. The primary objective is to improve energy efficiency. The prediction model is based on a cell division platform and is derived considering the additional flow during lane changing. An optimal control strategy is then developed, subject to the predicted trajectory generated for the preceding vehicle. Lane change prediction estimates future speed and gap of vehicles, based on predicted traffic states. The proposed framework outperforms the non-lane change traffic model, resulting in up to 13\% energy savings when lane changing is predicted 4--6 seconds in advance.
\end{abstract}


\maketitle

\section{Introduction}
The transportation sector plays a significant role in global energy demand, accounting for 60\% of the total oil consumption \cite{transportationfuelshare}. Additionally, it contributes to approximately 35\% of the total CO2 emissions \cite{transportationCO2share}. Recent advancements in connected autonomous vehicles (CAVs) have introduced remarkable opportunities for energy efficiency. CAVs are capable of perception and information broadcast. They utilize sensors to measure their own status as well as that of surrounding vehicles. Furthermore, communication occurs through vehicle-to-vehicle (V2V) and vehicle-to-infrastructure (V2I) connections, facilitating the exchange of critical information such as signal phasing and timing (SPaT) messages from connected signals. 

\begin{figure}
    \centering
    \includegraphics[width=0.6\textwidth]{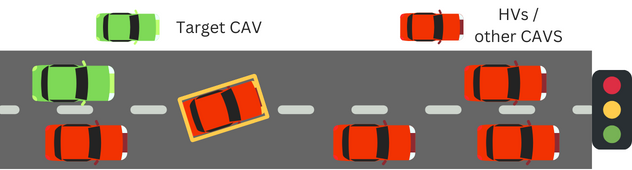}
    \caption{Problem description: other vehicles perform lane-changing in front of target CAV, which is inconsistent with the predicted driving plan}
    \label{fig:intro}
\end{figure}

The available level of automation in CAVs allows for precise computation and control. CAVs are capable of predicting traffic information, enabling them to plan future driving strategies, including optimal acceleration and deceleration patterns. By avoiding abrupt speed changes, an energy-efficient acceleration-deceleration strategy can be achieved, resulting in substantial reductions in energy consumption \cite{ShaoandSunb, Chalaki, hookeroptimal}. 
However, since the target CAV is aware of the traffic conditions ahead, it attempts to adjust its speed earlier than regular drivers. As a result, the target CAV reduces its speed gradually when approaching a red light or a congested zone. The target CAV maintains a slower speed than the leader vehicle, which increases the gap between them. As a result, the aforementioned speed control results in a higher lane change frequency in front of the controlled vehicle due to the relatively larger gap and lower speed. Unpredicted lane changes will result in additional energy consumption amounts for the target vehicle. Therefore, studying lane changes is crucial in order to propose the energy-efficient framework in more realistic cases when more than one lane is available. This study considers the impact of lane-changing on the predicted traffic states and consequently, on the optimal control plan when controlling a target CAV. Figure \ref{fig:intro} describes an example that will be addressed in this study. HVs or other CAVs may change their lane which is marked with a yellow box in the figure. If this lane change is close to and in front of the target CAV, it has a higher impact on the optimal control results. Lane changing is incorporated and predicted for either exiting or entering cases.

\subsection{Literature review}
Speed control is defined as a system that constantly and actively adjusts the speed. The very first speed controls are conventional cruise control and then adaptive cruise control \cite{marsden2001towards}. Advanced control platforms are presented afterwards such as implementing longitudinal models or various traffic models. Some studies employed model predictive control as a speed control framework such as focusing on V2V communication in \cite{geiger2012team} and using radar and signal information to predict optimal values for speed by a longitudinal model in \cite{asadi2010predictive}. Moreover, \cite{xu2018cooperative} used a longitudinal dynamic model to represent speed control with the goal of minimum travel time, and \cite{wang2019model} developed a similar approach with the goal of fuel efficiency. Many studies employed microscopic car-following models for the speed or acceleration control such as presenting various adaptive cruise control models for speed control \cite{goniros2019using}, employing Newell’s car-following model as the control algorithm for a cooperative adaptive cruise control \cite{shladover2012impacts}, demonstrating optimal velocity model to study the traffic flow \cite{geiger2012team}, exploring the cooperative control using intelligent driver model (IDM) \cite{milanes2014modeling}, and studying the impact of adaptive cruise controls on the traffic flow by microscopic modeling \cite{van2006impact}. On the other hand, some studies revealed that the macroscopic traffic modeling also results in a stable speed control mechanism employing a gas-kinetic (GKT) model in \cite{ngoduy2013instability} and \cite{delis2015macroscopic}, and a model based on the continuity equation in \cite{yi2006macroscopic}.

Among available models, cell transmission models (CTM) proposed by \cite{daganzo1994cell} are widely applied in the literature. Some studies employed CTM to manage and control a network, \cite{muralidharan2015computationally, carlson2010optimal} utilized ramp metering in order to control a freeway network. Many studies focused on speed limit controls; \cite{zhu2014accounting} used CTM in network modeling in order to control the speed limit dynamically, \cite{csikos2017variable} controlled the speed limit with the goal of maximizing network throughput, and \cite{han2017resolving, talebpour2013speed} employed CTM models to resolve freeway jam wave by controlling the speed limit. \cite{gomes2006optimal} demonstrated asymmetric cell transmission model to provide a ramp metering plan, \cite{lo2001dynamic} and \cite{alislam2020realtime} implemented CTM for traffic signal control. CTM models are also widely used to analyze and improve driving strategies, \cite{zhu2018modeling} employed the CTM model for speed adjustment and evaluated the corresponding emission and \cite{chen2020effects} modeled traffic flow in order to adjust the speed limit in real-time for autonomous vehicles with the goal of energy saving. There are also CTM-based models such as a discrete approximation of the Lighthill–Whitham–Richards (LWR) model which is a macroscopic traffic flow model in \cite{levin2016multiclass} and Payne’s (PW) second-order traffic flow model \cite{ecoapproach}. Recent studies have used the two aforementioned models for traffic prediction and providing optimal control based on that \cite{LWRpredict, ecoapproach}. \textcolor{blue}{ METANET simulator is also one example of employing such models for traffic modeling \cite{messmer1990metanet}.}

The aforementioned studies that employ CTM models are particularly for traffic analysis and prediction. The following step after traffic prediction is to present a framework to optimize the controlling strategy. There is a broad state-of-the-art in control strategies that demonstrate energy efficiency through connectivity and automation in vehicles \cite{zhang2019energy}. \cite{dong2022practical} proposed energy management strategies based on driving cycle predictions on a network scale. \cite{jiang2017eco, sun2020optimal} employed information from loop detectors and signal timing to minimize fuel consumption and emissions, demonstrating optimal speed control profiles for the target CAV. Furthermore, \cite{du2022comfortable} implemented reinforcement learning, and \cite{zhou2020stabilizing} focused on string stability, both presenting optimal speed control frameworks. Vehicle control can also be studied from a powertrain control perspective. \cite{wei2022co, liu2022bi, energyandmobility} has demonstrated the optimal control problem as a co-optimization involving vehicle dynamics, traffic constraints, and powertrain control.

Various prediction strategies have been demonstrated for CAV energy-efficient control, ranging from short-term to long-term platforms. Long-term platforms \cite{longtermpredict} are designed for traffic volume planning and traffic management rather than online control of a specific vehicle. However, controlling a CAV through a road section requires short-term predictions of future changes in traffic states. Our previous studies \cite{energyandmobility, ecoapproach} have shown the effectiveness of analytical traffic prediction for a prediction horizon of 10--15 seconds.

A large proportion of studies, such as \cite{wang100MPR, wu100MPR, hom100MPR} obtained their traffic states from 100\% CAV market penetration rate (MPR). However, the assumption of all vehicles being CAVs is not realistic in the expected future of this method \cite{WADUD}. This study presents a framework to control CAVs at various MPRs (mixed platoons). The presented traffic flow model is capable of predicting traffic information even with partial data of the entire platoon due to varying MPRs.

Among the macroscopic traffic models capable of short-term prediction, the first order LWR model can accurately describe traffic flow, stating ``flow equals density times speed'' \cite{treiberp81}, resulting in real-time traffic prediction \cite{LWRpredict}. However, speed in the first order models is directly dependent on the traffic density value, which leads to a lack of independent speed prediction, causing inaccuracy and fluctuations for short-term CAV control. Second-order models, such as the well-known Payne-Whitham (PW) model, are able to provide an independent and dynamic speed prediction, which is implemented in \cite{sun2021traffic}. Optimal control in coordination with an improved PW model presented 10--20\% energy benefits in our previous studies \cite{energyandmobility, techreport}.

\textcolor{blue}{In order to present a more realistic traffic study, particularly in the presence of CAVs, researchers model traffic flow in multiple-lane scenarios. As vehicles travel in different lanes in this problem, studying how vehicles change lanes to utilize all lanes of the road requires further consideration. The first research question in multi-lane sections is to model and develop how CAVs can perform LC such as developing a dynamic lane changing planning for CAVs by creating a cubic polynomial trajectory for LC maneuver in \cite{yang2018dynamic}} and explaining LC models and motion planning for CAVs and assessing the impacts of their LCs on traffic \cite{wang2019review}. \cite{du2020cooperative} presented a control framework to cover the lane changes of CAVs with the goal of safety. When more than one lane is available, control platforms are commonly implemented by car-following models along with lane-changing models \cite{wang2019review, sun2021cooperative, tao2004modeling, li2022simulation}, particularly in the presence of CAVs. \textcolor{blue}{ Another topic that received attention when mixed platoons utilize multiple-lane segments is how to assign lanes to CAVs with the goal of traffic throughput increase \cite{NAGALURSUBRAVETI2021103126} and efficiency \cite{kreidieh2022lane}.}

\textcolor{blue}{The second question in multi-lane sections is modeling the impact on traffic flow when vehicles change their lane. This practice particularly focuses on how follower vehicles will be affected when a vehicle enters their lane or leaves the lane. Early traffic modeling to include LC was performed by studying the interaction between adjacent lanes based on ``equilibrium density distribution'' with the goal of stability improvement in \cite{gazis1962density}. Studies began to derive the most critical parameters for predicting the number of lane changes in a lane  \cite{knoop2012quantifying, knoop2010lane}; they realized that density in the source and target lane were the key factors. \cite{roncoli2015traffic} modified a first-order, CTM-based flow conservation model to consider longitudinal and lateral flow for each lane that could be employed for considering the impact of LC. The goal of this study was to propose an optimal control for the flow in the presence of CAVs. Alternatively, car-following models can be modified to model the impact of LC on a follower vehicle such as modifying Newell's car-following model to analyze the behavior of the follower vehicle in \cite{chen2023modeling}. All mentioned models can be employed to predict LC in multi-lane sections but they are not able to anticipate the speed and location of a target CAV for a short-term prediction horizon. In other words, we cannot match these traffic models with the optimal speed control for a target vehicle due to prediction accuracy concerns.}

\cite{sun2021cooperative} presented an optimal LC model that is evaluated for only fully automated platoons and can be used for vehicle control. However, \cite{li2022simulation} employed the LC model in a mixed platoon to study the traffic performance and CAVs reaction. \textcolor{blue}{The latter studies have controlled LC for the target CAVs while the impact of lane changes performed by other vehicles when controlling a CAV has received less attention in previous studies. Optimal speed control strategy for CAVs heavily relies on traffic behavior in front of the CAV particularly when approaching a traffic light. As LCs change this behavior and also affect traffic speed, estimating those LCs will be crucial in maintaining energy benefits on multi-lane roads.}
Furthermore, there has been a lack of evaluation of a stable or safe framework for energy efficiency. The current study aims to address these two aforementioned concerns within the state-of-the-art.

\subsection{Contributions}
LC from adjacent lanes to the target lane \cite{LClaval, laval2004multi} is highly probable to occur immediately in front of the controlled CAV. The main reason is the control strategy tends to maintain a larger gap at the front, inducing LC events. Recent studies have worked on this phenomenon and considered the impact of LC maneuvers in speed control \cite{LCimpact1, SuiyiLC}. The CTM-based speed control frameworks rely on the prediction horizon however there is a lack of predicting lane-changing within this prediction horizon in the current studies. Therefore, the main concern still remains that upcoming lane changes could significantly impact the performance of the speed control, and \textcolor{blue}{no LC prediction has yet been both used in the traffic flow prediction and included in the speed control framework. Although various studies \cite{roncoli2015traffic, chen2023modeling, SuiyiLC} model the impact of LC on traffic flow, they are not accurate enough to be implemented in an integrated framework for controlling a target vehicle. In summary, this study first accounts for LC while predicting traffic states and second, modifies LC and traffic prediction to be matched with any optimal speed control and even vehicle powertrain models.}

Therefore, we propose a modified traffic prediction method capable of predicting lane-change events and assessing their impact on speed and density values. The main contributions of this work are as follows:

(1) We modify the PW model to incorporate the impact of LC on traffic state prediction.

(2) We implement a real-time LC prediction model that integrates with the modified PW model and utilizes data from vehicle connectivity. Furthermore, this study evaluates and calibrates the LC model to cover extensive scenarios.

(3) We examine the energy benefits by developing an optimal control framework capable of controlling target CAVs and introducing improved traffic constraints. After the implementations, the framework is able to record the impacts of the LC prediction method. 

(4) We conduct simulations of various scenarios and evaluate method robustness that includes: a network of two-lane signalized arterial corridors and traffic flow of mixed platoons for various MPRs and traffic volumes.

\textcolor{blue}{Although items (1), (2), and (3) have partially been explored in previous studies, this paper is the first to combine them into one control framework.}

The organization of the remainder of the paper is as follows. A general overview of the framework is introduced in Section \ref{frameworksec}. According to the framework, the first part is the traffic prediction, which corresponds to the LC prediction model and is discussed in Section \ref{trafficpredsec}. The second part explains the speed control in \ref{optimalcontrolsec}. Finally, the paper evaluates the quantitative performance and energy efficiency of the framework in Section \ref{resultssec}.

\section{Framework} \label{frameworksec}
\begin{figure}
    \centering
    \includegraphics[width=\textwidth]{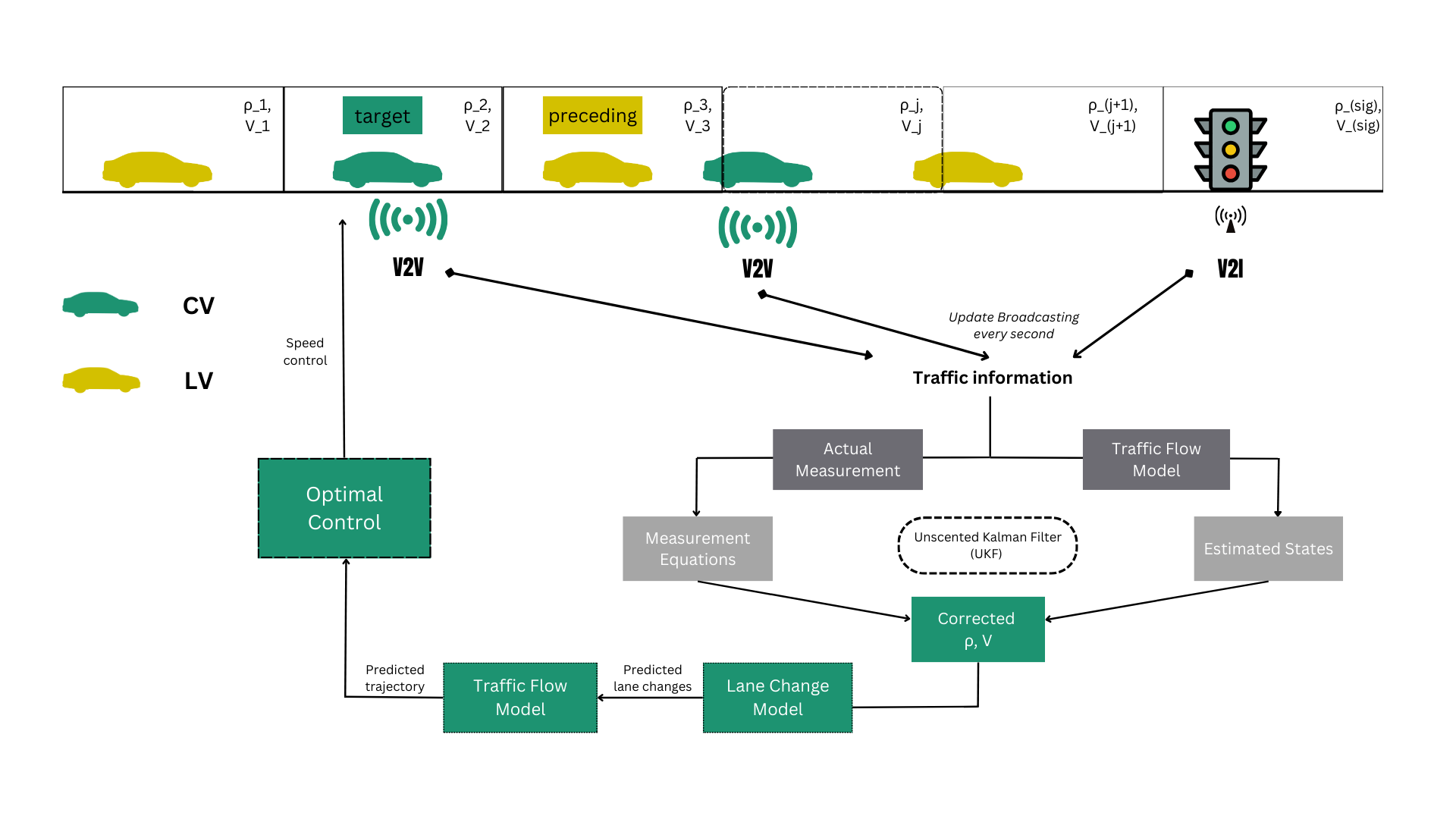}
    \caption{Framework demonstration and process including CAV and SPaT messages, UKF, traffic models, and finally the optimal control}
    \label{picframeworks}
\end{figure}

In this section, we explain how the platform works in a mixed platoon as shown as Fig. \ref{picframeworks}. CVs will provide traffic information including speed and location measurements in real-time. The traffic states need to be estimated from CV information using a state observer since the connected technology cannot measure and cover all the traffic stream. The traffic information is then applied in a cell-based traffic model to predict the traffic states in future. The traffic model is using a lane change model to also predict the impact of the lane changes on the traffic states. Given the future traffic states, we generate a framework to control a certain CV with the goal of optimal energy consumption.

In the first step, CVs measure and broadcast the data. CAVs can exchange data through V2V connections, encompassing speed and location information obtained from sensors, between CAVs and adjacent vehicles. Moreover, CVs are aware of future signal timing from signals equipped with V2I connections. Connections provide online traffic data exchange which is used to predict traffic states \cite{V2V}. Finally, the available data for our control includes the measured spacing and speed of the target CAV, the exchanged speed and distance values from other CAVs through V2V connections, and the future signal timing obtained via V2I connections.

The second step includes correction and estimation of traffic states via filtering. Only CVs can measure and transmit partial speed and spacing values on a segment. In other words, only a subset of the total vehicles on the road are connected vehicles (CVs) equipped with sensors and able to transmit the measure data. As a result, the precise location and speed information for every vehicle may not be readily available. Therefore, a state observer, an unscented Kalman filter (UKF), is implemented to provide an estimation for the traffic states \cite{mainUKF}. The UKF implementation is based on the previous studies, and readers are referred to \cite{suhUKF} for further explanation and implementation UKF on the traffic prediction model.

In the context of optimal control, the objective is to optimize the speed for the for the upcoming 10--15 seconds by a predictive control based on the predicted trajectory of a specific preceding vehicle. Although the optimal speed values are predicted for a horizon, only the value of the next time step is used to control the speed. However, a challenge arises when a new vehicle enters the space between the target CAV and the original preceding vehicle within this prediction horizon. The new vehicle changes the traffic states in both target lane and the source lane such as increasing the density of the target lane. Moreover, the new vehicle becomes the new preceding vehicle, for which the optimal control lacks a preexisting plan and predicted trajectory. Consequently, the optimal controller does not have the information of the new preceding vehicle along the 10--15 seconds horizon in advance. This disruption can negatively impact the optimal control. Instead, the platform capable of LC prediction, estimates that a lane change is anticipated to occur in the next few seconds, resulting in better traffic state prediction at an estimated location. With this enhanced knowledge, the optimal control system is afforded more time to make informed decisions, leading to reduced energy consumption during the process.

In the third step, the traffic prediction computes how traffic states including speed and density would change in the future for the next several seconds. In order to generate the traffic prediction, the road segment is divided into cells of equal length, and the model computes the density and speed of the next time step based on the current cell speed and density. Impact of the adjacent cells is also considered in the equations. LC prediction occurrence is synced with the traffic prediction model. To model LC, the traffic model will be updated considering the impact. Thus, The two aforementioned models work together to predict speed and density considering the impact of probable lane changes. 

In the final step of this process, the target CAV is controlled based on the previous steps. The target CAV utilizes measurements of the preceding vehicle's location at the current time step. Furthermore, the traffic flow model is propagated to estimate future cell densities and speeds. With these inputs, a trajectory is generated to predict the movement of the preceding vehicle within the next 10--15 seconds. To effectively manage the target CAV across the cells ahead, an optimal control mechanism is employed, aiming to minimize energy consumption. This mechanism computes speed values for the target CAV during the prediction horizon, facilitating precise speed control and ensuring optimal operation.

There are several assumptions made for the methodology that are explained as follows:

\textit{CAVs assumptions}: As aforementioned, we study mixed platoons of HVs and CVs in various proportions of connectivity. CAVs are assumed to be able to receive SPaT messages and information from other CAVs. CAVs can measure and transmit their own speed, location, spacing to the leader vehicle, and relative speed to the leader vehicle if applicable. Therefore, each CAV is able to provide information about two vehicles. This assumption is widely applied in prior studies \cite{energyandmobility, zhang2019energy, ecoapproach}.

\textit{Traffic behavior assumptions}: We assume that traffic follows normal driver behavior, meaning that vehicles drive based on typical car-following behavior and maintain a reasonable gap when following the leader vehicle. Based on this assumption, the proposed traffic models provide a valid approximation for predicting traffic states.

\textit{Lane change model assumptions}: This study focuses on predicting LCs motivated by speed gain which also has a higher impact on the speed control. Other reasons for the lane changing such as lane changes due to routing can be added to the traffic model. However, they are not predicted by the lane change model since they mostly require network traffic assignment which is not in the scope of this study. it is assumed that there is no requirement for vehicles to exclusively drive in the rightmost lanes or keep the leftmost lane clear, and all lanes have equal priority for vehicles when traveling at free-flow speeds. 

\textit{Optimal control assumptions}: The optimal control employs the vehicle power model to control the vehicle longitudinally. Moreover, the control is based on the vehicle acceleration, which is directly related to speed, but the vehicle powertrain level is not studied. The main reason is that the current study focuses on the modified traffic model, along with lane-change prediction to evaluate performance, not presenting a novel vehicle control strategy.

\section{Traffic Prediction} \label{trafficpredsec}
This section presents a modified traffic prediction model to consider the impact of lane changing phenomena on the traffic states. The lane-change process is characterized by the transfer of flow between lanes. This flow transfer has a direct impact on the road segment, resulting in changes in density and speed within that segment. The impact is derived and added to the PW model in order to consider the additional flow. PW model is a second order traffic flow model that describes density and speed independently for the next time step \cite{payne1979}. In the current time step, estimated traffic states are available from the estimation step. The road segment is divided into shorter cells, and the traffic flow model will be propagated to predict the density and speed of the cells for the cells for the next time steps. This process is being done for every cell in the broadcast range of the target CV in order to have the cell states during the prediction horizon. Each cell is microscopically large to have vehicles in it and macroscopically small to be able to capture the changes in factors that impact the flow. 

As mentioned, the second order traffic flow model has two terms to describe density and speed separately. In this part, we derive the equations from the kinematic wave theory and obtain a macroscopic equation for the density and speed. When LC does not occur, the equations are equal to the standard PW model since there is no additional flow entering or exiting between the lanes. 

Due to the varying lane-change flows across different lanes, the modified model is propagated separately for each lane. The sign of the LC flow differs per lane, with a positive value for the target lane and a negative value for the source lane. Consequently, the LC term in the equations assumes different values for each lane. 

The cell length and time step should be relatively short for the cell-based traffic model in order to satisfy some assumptions: lane-changing flow is assumed to be uniformly distributed in the length of a cell. Furthermore, the speed of lane changing vehicle is assumed to be equal to the adjacent cell's speed. In the case of longer cell length, a higher number of vehicles can fit in each cell, and as a result, the cell states are not approximated to the lane-changing flow only.

\subsection{Density equation derivation}
From the hydrodynamic flow-density relationship, we can model the traffic stream in a cell of the road. We first assume the cell is homogeneous, without any additional flow from other lanes, which implies no lane changes. The density equation will be further derived to incorporate the lane-change events.
We start from the continuity equations which explains the change of the flow in the cell.

The number of vehicles for a road cell with length of $ \Delta x$ at time t is given by \cite{treiberbook}:
\begin{equation}
   n(t) =  \int_{x}^{x+\Delta x} \rho(x',t) \,dx'  \approx \rho(x,t) \Delta x
\end{equation}
where $x$ and $x+ \Delta x$ are the start and end points of the cell and $\rho(x',t)$ is the density at location $dx'$ and time $t$.  

For the rate of change in the number of vehicles, we can write:
\begin{equation}
    \label{eqdn1}
    \frac{dn}{dt} = \frac{\partial \rho}{\partial t} \cdot \Delta x
\end{equation}
This equation would be available since $\Delta x$ (road cell length) is constant with respect to time.

\begin{figure}
    \centering
    \includegraphics[scale=0.21]{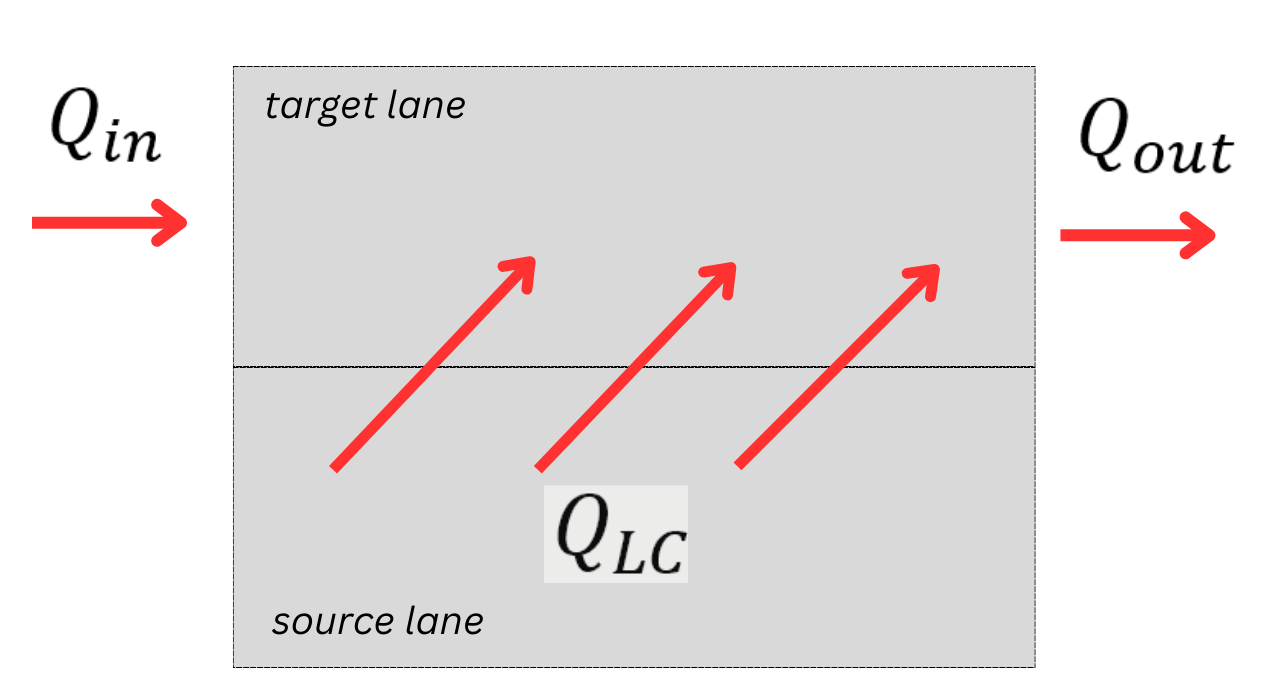}
    \caption{Sketch of the incoming and outgoing flows for the target cell. In this case, additional flow will be entered to the target cell from the source lane due to LC in addition to upstream and downstream flows}
    \label{picflows}
\end{figure}

For each cell, assuming no nonuniform events occur within the middle of the cell, there is only incoming flow and outgoing flow. Given the assumption of a homogeneous road cell, alterations to the number of vehicles can only be attributed to the inflow $Q_{\mathrm{in}}$ or outflow $Q_{\mathrm{out}}$ at the cell boundaries:
\begin{equation}
    \frac{dn}{dt} = Q_{\mathrm{in}}(t) - Q_{\mathrm{out}}(t)
\end{equation}
The flow could be presented as $Q_{\mathrm{tot}}$ which is a function of the time and location. Upstream and downstream flows are analyzed at the same time, but at two different locations:
\begin{equation}
    \label{eqdn2}
    \frac{dn}{dt} = Q_{\mathrm{tot}}(x,t) - Q_{\mathrm{tot}}(x+\Delta x, t)
\end{equation}
Equations \eqref{eqdn1} and \eqref{eqdn2} are equal in the left side, so, we can write the change in the flow from upstream to downstream as:
\begin{equation}
    \label{eqrho1}
    \frac{\partial \rho }{\partial t} = - \frac{Q_{\mathrm{tot}}(x+\Delta x, t) - Q_{\mathrm{tot}}(x,t)}{\Delta x}
    \approx \frac{\partial Q_{\mathrm{tot}}(x,t)}{\partial x}
\end{equation}
Finally, using the hydrodynamic flow-speed relationship $Q_{\mathrm{tot}} = \rho V$: 
\begin{equation}
    \frac{\partial \rho}{\partial t} + \frac{\partial (\rho V)}{\partial x} = 0
\end{equation}
Now, we need to incorporate the lane changing phenomena. In addition to the inflow $Q_{\mathrm{in}}$ and outflow $Q_{\mathrm{out}}$ at the cell boundaries, changes in entering or exiting flow from adjacent lanes due to lane-change can also affect the number of vehicles. This additional flow will be included in the calculation of the number of vehicles in the cell shown in Fig. \ref{picflows}. Consequently, Equation \eqref{eqdn2} needs to be revised as follows:
\begin{equation}
    \frac{dn}{dt} = Q_{\mathrm{tot}}(x,t) - Q_{\mathrm{tot}}(x+\Delta x, t) + Q_{\text{LC}}(x,t)
    \label{eqQtot}
\end{equation}
where $Q_{\text{LC}}(x,t)$ represents the flow of vehicles that change lanes at the given cell during time $t$. If vehicles are entering the target lane, the flow values will be positive, but if they are exiting the target lane, the flow values will be negative. Since Equation \eqref{eqQtot} is derived from the hydrodynamic flow equation, $Q_{\text{LC}}$ can take on any value (provided that the cell has sufficient capacity to accommodate it). If one vehicle is entering the cell, $Q_{\text{LC}}$ will be 1 vehicle over the time step. The only requirement for $Q_{\text{LC}}$ is that the flows resulting from lane changes are constant along the length $\Delta x$ \cite{treiberbook}. This assumption holds for a sufficiently short cell length.
Similar to Equation \eqref{eqrho1}, for the density change with respect to the time we have:
\begin{equation}
    \frac{\partial \rho }{\partial t} = - \frac{Q_{\mathrm{tot}}(x+\Delta x, t) - Q_{\mathrm{tot}}(x,t)}{\Delta x} + \frac{Q_{\text{LC}}(x,t)}{\Delta x}
    \label{added_QLC}
\end{equation}
\begin{equation}
    \frac{\partial \rho}{\partial t} + \frac{\partial (\rho_{\mathrm{tot}}V)}{dx} = \frac{Q_{\text{LC}}(x,t)}{\Delta x}
\end{equation}
where $Q_{\text{LC}}(x,t) = \rho_{\text{LC}}(x,t) \cdot V_{\text{LC}}(x,t)$, and $V_{\text{LC}}(x, t)$ is the speed of the vehicles in the adjacent lane that makes the lane change. $\rho_{\text{LC}}(x,t)$ is the lane change density along the road cell (i.e., the number of vehicles changing lanes over the length of the cell, which can also be fractional). Since the lane change flow is assumed to be uniformly distributed, the value of $\rho_{\text{LC}}(x,t)$ is considered constant within each cell. In other words, the number of vehicles changing lanes remains unchanged when evaluating each cell for a relatively short period of the time. The time interval should be shorter than the LC phenomena duration in order to make the assumption valid. An estimation for the density value will be presented in the next subsection. Here, $\Delta x$ represents the length of the cell.

The density equation can be expressed in a discrete form now, considering each $dt$ and $dx$.
\begin{equation}
    \frac{\rho_{j}(t+1) - \rho_{j}(t)}{\Delta t} + \frac{ \rho_{j}(t) \cdot V_{j}(t) - \rho_{j-1}(t) \cdot V_{j-1}(t)}{\Delta x} = \frac{Q_{j,\text{LC}}(x,t) }{\Delta x}
\end{equation}
In the cell coordinates, we have cell $j$ corresponding to location $x$ and cell $j-1$ stands for location $x-\Delta x$, representing $\Delta t$ and $\Delta x$ as time step and cell length respectively. Thus, the future density of cell $j$ at the next time step is:
\begin{equation}
    \rho_{j}(t+1) =  \rho_{j}(t) - \frac{\Delta t}{\Delta x} \cdot \left( \rho_{j}(t) \cdot V_{j}(t) - \rho_{j-1}(t) \cdot V_{j-1}(t) \right) + \frac{\Delta t}{\Delta x} \cdot \rho_{j,\text{LC}}(t) V_{j,\text{LC}}(t)
    \label{eqdensitycell}
\end{equation}
\textcolor{blue}{where $\rho_{j,\text{LC}}$ is defined as LC density that has interaction with cell $j$ (exiting or entering target lane, occurring inside cell $j$), and similarly $V_{j,\text{LC}}$ is the speed of LC vehicles within location of cell $j$ (exiting or entering target lane).}

In cases where complete information about all available vehicles and their lateral locations is available, we can record the value of $\rho_{\text{LC}}$, which represents the number of vehicles switching lanes per cell length in kilometers. However, in general, computing $\rho_{j,\text{LC}}(t)$ would require constant monitoring of each lane, which can be challenging and involve additional parameters. Moreover, monitoring vehicles is only able to provide the current time step lane change density not any estimation of $\rho_{j,\text{LC}}(t)$ for the prediction horizon. To address realistic problems and leverage the capabilities of connected vehicles, we provide an estimate of the lane changing density several time steps in advance based on certain assumptions. \textcolor{blue}{Thus, the value of $V_{j,\text{LC}}$ can also be defined as the average speed of LC vehicles. For example, if one vehicle is estimated to change the lane within the location of cell $j$, $V_{j,\text{LC}}$ is equal to the speed of that vehicle only. If the lane changing vehicle's speed is available through connectivity, the exact value will used for $V_{j,\text{LC}}$. However, if the speed of that vehicle is not accurate (e.g. that vehicle is not CAV, neither measurement is available for that vehicle), we use the adjacent cell speed value as an estimation. Having a short cell length ensures relatively low errors in this approximation.}

\subsubsection*{Lane change density parameter} \label{LCdensitysec}
In this study, we propose an estimation for $\rho_{j,\text{LC}}(t)$ that has been validated using Simulation of Urban MObility (SUMO) simulations and calibrated for various scenarios. Through literature review and experiments, it has been found that the lane changing density depends on several factors in the simulation, such as the speed ratio between the target lane and the adjacent lane at a given location, the density difference of neighboring roads, and the actual density of each lane.

However, it is important to note that within the scope of this study, information regarding driver decision-making and behavior is not available to accurately capture the lane change density. The only data available is the estimated cell density and speed for each lane, which is provided by CVs. Therefore, our approach focuses on predicting the density value from the available cell states. To achieve this, we have implemented a lane change model that estimates the probability of a lane change occurring at a given cell.
\begin{equation}
    \rho_{j,\text{LC}}(t) = \frac{I(V,\rho) \cdot \alpha}{\Delta x}
\end{equation}
In the proposed estimation, the indicator function $I(V,\rho)$ is used to estimate the event of a lane change occurring. We introduce a modification factor, $\alpha$, which adjusts the occurrence of lane changes within each time step. This modification factor takes into account that lane changing phenomena may take longer or shorter than the time step duration ($\Delta t$). By adjusting $\alpha$, we can control the impact of lane changes in relation to the time step. In summary, $\frac{1}{\alpha}$ is the duration of the lane change. The derivations are originally based on how the density changes with respect to time. In hydrodynamic flow, the specific rate of density change could be any continuous value, not necessarily a discrete number. However, for our estimation, we consider a discrete representation of lane changing phenomena to capture it as individual vehicles. As a result, the estimated LC event indicator only outputs values from the set \{-1, 0, 1\} meaning one or zero vehicles. Since time step and the cell length in the traffic flow model is short enough to cover a single lane-change at a time, it is reasonable to assume at most one lane change can happen for a single cell at each time step. For instance, if we are interested in observing the time it takes for a vehicle to change lanes, it has been mentioned in \cite{timeLC} that the total duration of a lane change is approximately 1 s. However, our estimation aims to capture a full lane change. Therefore, if the defined time step is not sufficient to observe a lane change within the expected duration or longer, the estimated $I(V,\rho)$ needs to be modified using the modification factor $\alpha$. To calculate $\alpha$, if the time step is less than 1 s, it will be decreased to provide a fractional density for that cell. Conversely, if the time step is greater than 1 s, $\alpha$ will be increased to indicate a higher value for the lane change density, particularly in those intervals where lane changes are highly probable. For example, if a vehicle enters the cell, the estimated probability should be equal to 1. Finally, $\Delta x$ is the cell length. For each cell we may have these three cases:
\begin{equation}
     \rho_{j,\text{LC}}(t) =
    \begin{cases}
        \frac{\alpha}{\Delta x} & \text{1 vehicle enters target lane, $I(V,\rho)=1$} \\
        0 & \text{no LC happens, $I(V,\rho) = 0$} \\
        \frac{-\alpha}{\Delta x} & \text{1 vehicle leaves target lane, $I(V,\rho)=-1$} 
    \end{cases} 
    \label{rhoLCestimation}
\end{equation}
In the modified PW model definition, we make the convention that if more than half of the width of a vehicle is inside a cell in a certain lane, the entire vehicle's density belongs to that cell. Conversely, if less than half of the vehicle's width is inside an adjacent cell in the adjacent lane, it is assumed to be an empty cell. Although, since both source and target lanes are still occupied by the lane-change vehicle and no other vehicle can be at that location, the framework is programmed to block either lanes at the location due to LC occurrence. Additionally, the defined time step for the modeling in this study represents a digitized view is required.

However, for any two adjacent cells in the same lane, the framework calculates fractional densities to accurately capture the vehicles that are at the boundary between the cells. This allows for a more precise representation of the distribution of vehicles across adjacent cells among the same lane.

\subsection{Speed equation derivation}
In macroscopic first-order models, such as the Lighthill-Whitham-Richards (LWR) model, the local speed and flow are statistically coupled to the density through the fundamental diagram. However, in second-order models, the local speed has an independent dynamic acceleration equation that describes speed changes as a function of density, local speed, their gradients, and other external factors. In the speed equation, the focus is on the driver's perspective, which is why the local speed is used. This means that the rate of speed change is derived based on the trajectory (Lagrangian coordinates) rather than a stationary point, taking into account the changing location over time. Moreover, the equilibrium speed is computed for the traffic flow model as a function of speed and density. In time-continuous second-order models, the acceleration equation takes the form of a second partial differential equation, as described in \cite{treiberbook}:
\begin{equation}
    \frac{dV(x,t)}{dt} = \frac{\partial V}{\partial t} + V(x,t) \frac{\partial V}{\partial x} = A[\rho(x,t),V(x,t)]
\end{equation}
where $V(x,t)$ is the speed in location $x$ at time $t$ and $A$ is the acceleration which is function of the speed and density at a point. \cite{treiberbook} mentions that the acceleration includes these terms:

1. Mean acceleration of the vehicles in the neighbourhood where they want to reach the local steady state or equilibrium speed $(V_e)$. The time required to reach the equilibrium speed is called the speed adoption time ($\tau$) which is assumed to be constant. So the acceleration would be:
\begin{equation}
    A_{\mathrm{relaxation}}(x,t) = \frac{V_e(\rho(x_a,t)) - V(x,t)}{\tau}
\end{equation}
2. The collective response of nearby vehicles also influences acceleration \cite{treiberbook134138}, leading to changes in density and speed gradients. This effect arises from the kinematic impact of speed variations within the traffic flow and also the driver reactions, commonly referred to as traffic pressure. For instance, when transitioning from a lower density region at the beginning of a  to a higher density area at the ending locations, it results in slower-moving vehicles downstream. As more vehicles with lower speeds accumulate downstream, the local macroscopic speed of the entire  decreases. The speed variation is quantified as the density change relative to the location, considering the speed variance, as discussed in \cite{treiberbook134138}:

\begin{equation}
    A_{\mathrm{kin}}(x,t) = -\frac{c_0^2}{\rho(x,t)} \cdot \frac{\partial \rho}{\partial x}
\end{equation}
where $c_0^2$ is the traffic pressure constant which represents the speed variance. According to \cite{treiberbook134138}, it is valid to assume that the speed variance is constant.

3. Diffusion: diffusion describes the distribution of the vehicle densities \cite{diffusionreference}. In this study, it is assumed that there is no diffusion due to the short cell length and using the Lagrangian coordinates for the speed as mentioned in \cite{diffusionreference2}.

4. Merging and diverging in the target lane. This term is available when we have $V_{\mathrm{target\text{-}lane}} > V_{\mathrm{adjacent\textbf{-}lane}} $. The total derivative $\frac{dV}{dt}$ of the local speed denotes the rate of average speed change of vehicles in a small road with length $\Delta x$. Similar to the density equation, it is assumed that additional density due to LC is uniformly distributed over this length. Thus, there is a kinematic acceleration on behalf of speed change in elements available on the cell; which is assumed to be constant during each time step. In other words, the speed from the actual value of $v_{\mathrm{original}}$ in upstream will be reduced to $V_{\mathrm{adjacent\textbf{-}lane}}$ in the downstream. The partial derivative would be included in rate of changing number of vehicles and also rate of speed change. The rate of cell's speed change is the average of speed change for all $n$ vehicles in the road cell ($n = \rho \Delta x$):
\begin{equation} \label{averagespdeq}
   \frac{dV}{dt} = \frac{d}{dt} \left(\frac{1}{n} \sum_{\alpha=1}^{n} v_{l} \right)
\end{equation}
where $v_{l}$ is the speed of each vehicle in the cell.
The average speed change in Equation \eqref{averagespdeq} includes the derivative of two terms that can be written separately: \textcolor{blue}{1) the partial derivative with respect to the change in the number and 2) the partial derivative with respect to the speed change for all vehicles in the cell.} 

First, suppose that LC is the sole reason for the change in the number of vehicles within the cell. Like the change in the density in Equation \eqref{added_QLC} from lane changing, additional flow from LC explains the rate of the change in the number of vehicles \cite{treiberbook}:
\begin{equation} \label{eq:dn/dt}
   \frac{dn}{dt} = Q_{\text{LC}}
\end{equation}
Second, regarding the speed change of all $n$ vehicles, it is important to note that vehicles previously in the lane have no impact on the rate of the cell's speed change. Only new LC vehicles will influence the speed function, and thus, the change in total speed is solely attributed to these new LC contributors among all vehicles present in the cell at the end of the time step.:
\begin{equation} \label{eq:dv/dt}
   \frac{d}{dt} \left( \sum_{\alpha=1}^{n} v_{l} \right) = Q_{\text{LC}} V_{\text{LC}}
\end{equation}
Finally, the acceleration term that is only related to LC will be the partial derivative of the combination of two functions:

\begin{equation} \label{eq:LCacceq0} {\color{blue}
        A_\text{LC}(x,t) = \frac{d}{dt} \left( \frac{1}{n(t)} \sum_{\alpha=1}^{n} v_{\alpha} \right)\\ = \frac{\partial \frac{1}{n(t)}}{\partial t} \cdot \left( \sum_{\alpha=1}^{n} v_{l} \right) + \frac{\partial \left( \sum_{\alpha=1}^{n} v_{l} \right)}{\partial t} \cdot \frac{1}{n}   
        }
\end{equation}

\textcolor{blue}{Combine Equations \eqref{eq:dn/dt}, \eqref{eq:dv/dt}, and \eqref{eq:LCacceq0} to obtain lane-change acceleration equation}
\begin{equation} \label{LCacceq}
    A_\text{LC}(x,t) = - \frac{Q_{\text{LC}} V}{n} + \frac{Q_{\text{LC}}V_{\text{LC}}}{n} = \frac{Q_{\text{LC}}(V_{\text{LC}} - V_x(x,t))}{\rho \Delta x}
\end{equation}

It is important to note that Equation \eqref{LCacceq} only represents the speed change on behalf of LC vehicles. The overall speed change depends on all mentioned terms which is summarized as follows. Overall, the rate of speed change for the cell includes multiple acceleration terms:
\begin{equation}
    \frac{\partial V}{\partial t} + V(x,t) \frac{\partial V}{\partial x} = \frac{V_e(\rho(x,t))  - V(x,t)}{\tau} - \frac{c_0^2}{\rho} \cdot \frac{\partial \rho}{\partial x} + A_{\text{LC}}
\end{equation}
For the cell division in discrete time steps similar to definitions for Equation \eqref{eqdensitycell}, $\Delta t$ and $\Delta x$ can be written as the time step and cell length for short interval values. The traffic pressure term also considers the density of one cell ahead. 
\begin{equation}
    \frac{V_x(t+1) - V_x(t)}{\Delta t} + V(x,t) \frac{V_{x}(t) - V_{x-1}(t)}{\Delta x} = \frac{V_e(\rho(x,t)) - V_x(t)}{\tau} - \frac{c_0^2}{\rho_x(t)} \cdot \frac{\rho_{x+1}(t) - \rho_x(t)}{\Delta x} +  \frac{ Q_{\text{LC}}(V_{\text{LC}} - V_x(t)}{\rho_x(t) \Delta x}
    \label{eqvpwcontinous}
\end{equation}
 For cell discretization, cell $j$ represents location $x$ and cell $j-1$ represents location $x$. In discrete time and space we have $dt$ and $dx$. Also, we have $Q_{\text{LC}} = \rho_{\text{LC}} V_{\text{LC}}$ and we can use the presented density function for $\rho_{j,\text{LC}}(t)$. A small value $\epsilon$ is added to avoid a zero denominator.
\begin{equation}
\begin{aligned} 
    V_j(t+1) = V_j(t) - \frac{\Delta t}{\Delta x} V_j(t) [V_{j}(t) - V_{j-1}(t)] + \Delta t \cdot \frac{V_e(\rho(x,t)) - V_j(t)}{\tau} - \frac{\Delta t} {\Delta x} \cdot c_0^2 \frac{\rho_{j+1}(t) - \rho_j(t)}{\rho_j(t) + \epsilon}\\ 
    + \frac{ \rho_{j,\text{LC}}(t) V_{j,\text{LC}} (t) ( V_{j,\text{LC}} - V_j(t))}{\rho_j(t) + \epsilon}  \frac{\Delta t}{\Delta x}
    \label{eqvpw}
\end{aligned}
\end{equation}

\subsection{Modified PW model}
As a summary of derivations, we present a second order model based on the PW model. Traffic density and speed will be predicted for the next time step via density and speed equations. This process is propagated for the prediction horizon to facilitate the optimal control. The traffic states for cells are predicted via the modified PW model:
\begin{equation}
        \rho_{j}(t+1) =  \rho_{j}(t) - \frac{\Delta t}{\Delta x} \cdot \left( \rho_{j}(t) \cdot V_{j}(t) - \rho_{j-1}(t) \cdot V_{j-1}(t) \right) + \frac{\Delta t}{\Delta x} \cdot \rho_{j,\text{LC}}(t) V_{j,\text{LC}}(t)
        \label{eqmodifyrho}
\end{equation}

\begin{equation}
\begin{aligned}    
        V_j(t+1) = V_j(t) - \frac{\Delta t}{\Delta x} V_j(t) [V_{j}(t) - V_{j-1}(t)] + \Delta t \cdot \frac{V_e(\rho(x,t)) - V_j(t)}{\tau} - \frac{\Delta t} {\Delta x} \cdot c_0^2 \frac{\rho_{j+1}(t) - \rho_j(t)}{\rho_j(t) + \epsilon}\\
        + \frac{ \rho_{j,\text{LC}}(t) V_{j,\text{LC}} (t) ( V_{j,\text{LC}} - V_j(t))}{\rho_j(t) + \epsilon}  \frac{\Delta t}{\Delta x}
        \label{eqmodifyv}
\end{aligned}
\end{equation}

Equations \eqref{eqmodifyrho} and \eqref{eqmodifyv} are conservation and acceleration equations from the PW model considering the lane change flow in discrete system. The equilibrium speed $V_e(\rho(x,t))$ is computed as follows \cite{energyandmobility}:
\begin{equation}
     V_e(\rho(x,t)) =
    \begin{cases}
        v_0,  & 0 \le \rho_j(k) \le \rho_c \\
        
        c \cdot (\rho_{\mathrm{jam}} / \rho_j(k) - 1),   & \rho_c \le \rho_j(k) \le \rho_{\mathrm{jam}}
    \label{eqVe}
    \end{cases} 
\end{equation}

\begin{equation}
    \rho_c = \frac{\rho_{\mathrm{jam}}}{v_0/c + 1}
\end{equation}
where $v_0$ is free flow speed, $\rho_{\mathrm{jam}}$ is the jam density, $c$ is the slope of density drop in the density-speed chart between the critical density and the jam density, and $\rho_c$ is the critical density.

In the context of signalized intersections within the framework, it is important to consider that when a signal is in the red phase, the cell speed at the location of the signal is zero as vehicles are required to come to a stop. For simplification purposes, the yellow phase is also considered as part of the green phase since vehicle can still enter the intersection, resulting in two distinct phases for the signal: the green phase and the red phase. Additionally, an enhanced equilibrium speed algorithm is implemented to enhance the accuracy of predictions based on \cite{energyandmobility}. This algorithm takes into account the influence of the red light on the last several cells leading up to the signal. This modification aims to more accurately reflect the impact of the red light on traffic flow. Given the V2I connection, CVs are aware of the signal timing and the location of the intersection.

\subsection{Measurement Equations}
The distance between vehicles is closely related to the cell density and the vehicle speed is also related to the cell speed. Among all vehicles, CAVs can measure part of distance and speed values. Therefore, measurement equations can describe the relation between cell states and vehicle information via linear interpolation as explained in \cite{suhUKF}.
\begin{equation}
    \label{eqvmeasure}
    y_{v_i}(k) = \lambda_i(k) v_{{j_i}+1}(k) + [1 - \lambda_i(k)] v_{j_i}(k) + w_{v_j} 
\end{equation}
\begin{equation}
    \label{eqrhomeasure}
    y_{d_i}(k) = \beta_i(k) \frac{1}{\rho_{{j_i}+1}(k)} +  [1 - \beta_i(k)] \frac{1}{\rho_{j_i}(k)} + w_{d_j} 
\end{equation}
where $y_{v_i}$ is the speed of the $i$-th captured vehicle via V2V, including both CVs and their adjacent 
vehicles; $j_i$ is the index of the last cell the preceding vehicle 
passed; $\lambda_i(k) = d_{vi}(k)/dx - j_i$ is the interpolation 
coefficient; $d_{vi}(k)$ is the location of the $i$-th preceding CV; 
$w_{v_j} $ is a Gaussian random variable to model measurement 
uncertainties. $ y_{d_i}$ is the car-following distance of two consecutive 
vehicles; $\beta_i(k) = d_{di}(k) /dx - j_i$ is the interpolation coefficients; $ d_{di}(k)$ is the middle position between two consecutive vehicles; $j_i$ is the index of the left cell near $d_{di}(k)$, and $ \rho_{j_i}$ is the density of the cell $j_i$. Readers are referred to \cite{suhUKF} for demonstration details.

\subsection{Lane change model}
This study on CV control focuses on predicting LCs motivated by speed gain. This section and proposed analysis is conducted based on these speed-gaining lane changes. In other words, if a faster lane is available, there is a motivation to change lanes; otherwise, lane changes are not predicted to occur. The reason for this type of lane changing is that the target vehicle vehicle is trying to save energy and
as a result creates a larger gap; as a result, other vehicles may decide to use that
gap in order to increase their speed. It should be noted that other types of lane changes, which may involve specific geometric considerations or vehicle routing strategies (e.g., lane changes for left or right turns), are not predicted by this model. If other motivations for lane changing can be predicted, then the modified traffic model is capable of predicting the corresponding traffic states. Furthermore, it is assumed that there is no requirement for vehicles to exclusively drive in the rightmost lanes or keep the leftmost lane clear, and all lanes have equal priority for vehicles when traveling at free-flow speeds.

This section presents an attempt to predict lane changes to update the traffic flow model using an analytical model. To address realistic problems and leverage the capabilities of CVs, we can provide an estimation of the LC density ($\rho_{\text{LC}}$) in Equations \eqref{eqmodifyrho} and \eqref{eqmodifyv} several time steps in advance. The modified traffic flow model incorporates non-zero lane change density to account for lane change occurrences. A prediction method is developed and integrated into the cell model to determine when and where lane changes take place. The presented method is adaptations of the Krajzewicz lane-changing model, tailored to the available traffic state information \cite{LCmodel}. This study utilizes cell occupancy data to detect vehicles and estimate the probability of lane changes at each cell. Where a CV is available, the measured speeds and distances are applied to enhance the model accuracy. 

In the Krajzewicz lane-change (LC) model, each vehicle compares the feasible speed in its current lane with the feasible speed in the adjacent lane at the closest $x$ coordinate. If there is a significant difference between these two values, the driver attempts to make a lateral movement (a lane change). The model introduces a benefit function that can be computed at each time step and added to the cumulative values for each vehicle from previous steps. This benefit function serves as a criterion for determining the desirability of a lane change for a given vehicle. A positive value represents a benefit for changing lanes from the adjacent lane to the target lane due to a speed increase.
\begin{equation}
    \label{eqbenefit}
    b_{l_t}(t) = \frac{v_{\mathrm{safe}}(t,l_t) - v_{\mathrm{safe}}(t,l_c)}{v_{\max}(l)}
\end{equation}
where $b_{l_t}(t)$ is the benefit of a vehicle to change to a target lane $l_t$ at time $t$, $l_c$ and $l_t$ are the vehicle’s current and target lanes, respectively, $v_{\mathrm{safe}}(t,l)$ is the speed the vehicle could drive safely with on lane $l$ at time $t$
(in m/s), and $v_{\max}(l)$ is the maximum speed the vehicle can take which assumed to be free flow speed. The values of safe speed for each lane are calculated as follows \cite{LCmodel}:
\begin{equation}
    \label{eqsafespeed}
    v_{\mathrm{safe}}(t) = \min \left(v_{\max}(l), -\tau \cdot b + \sqrt{(\tau \cdot b)^2 + v_{\mathrm{leader}}(t-1)^2 + 2 \cdot b \cdot g(t-1) } \right)
\end{equation}
where $v_{\mathrm{safe}}(t)$ is the safe speed at time t; $\tau$ is vehicle's reaction time, $b$ is the maximum deceleration ability of the vehicle, $v_{\mathrm{leader}}(t)$ is speed of the leader, and $g(t)$ is the gap between the vehicle and the leader (bumper to bumper) at time $t$. Finally, the safe speed must be less than or equal to the free flow speed. In other words, when there is a large gap, vehicles are driving with the free flow speed rather than the value computed by the right part of Equation \eqref{eqsafespeed}.

As mentioned, the model takes into account the quantitative benefit that is associated with the potential speed increase due to lane change. The value of benefit at a given time step is computed, and the corresponding value is cumulatively added to a variable called benefit memory comprehensively explained in \cite{LCmodel}. The benefit memory models the magnitude of LC motivation for a single vehicle quantitatively. The main reason for selecting the cumulative benefit memory concept is to model the driver behavior by quantifying the speed gain from the current time step up to several seconds in advance. This process is done for every single cell or every single vehicle at each time step.  However, if the current lane is already faster than the adjacent lane (indicating no benefit for lane changing), the cumulative memory value is divided by two. The memory value for each vehicle is updated at each time step. When it exceeds a predefined threshold, the model predicts that a lane change will occur for that specific vehicle at that time step. The impact of the threshold value will be presented in the results section.

It is important to note that for a lane change to occur, there must be sufficient space available for the lane-changing vehicle, which is assumed greater or equal to the length of the vehicle plus two minimum gaps (one for the front, one for the rear). \textcolor{blue}{However, if the predicted gap is less than this amount (the length of the vehicle plus two minimum gaps) at a moment, the vehicle will delay the lane change until the gap becomes sufficient. At that time, the vehicle changes the lane if the benefit from equation \eqref{eqbenefit} is still exceeding the pre-defined threshold (In the results section it will be discussed that values in the range 2--2.5 are reasonable).}

A numerical example will be presented to clarify this process. Assuming there are two lanes, the free flow speed for both lanes is 20 m/s. There is a vehicle in lane 1 that faces a slower preceding vehicle cruising at a constant speed of 15 m/s. In the previous time step, the gap between the two vehicles was measured to be 15 m. Using Equation \eqref{eqsafespeed}, the safe speed for the mentioned lane is around 15.35 m/s. However, the safe speed in adjacent lane is still free flow speed (20 m/s). The benefit at this time step would be:
\begin{equation}
    b_{l_n}(t) = \frac{20 - 15.350}{20} = 0.233
\end{equation}

This value will be added to the prior benefit memory, and when it exceeds the threshold of 2 in this example, the vehicle will decide to perform the lane change. Choosing shorter time steps will result in better lane change predictions. The value of threshold should be determined for the scenario. In the results section, the various threshold values are examined.

\subsubsection*{LC model for cell occupancy estimation}
Being aware of the current location of all vehicles with their respective lane numbers can provide a simple and accurate lane change prediction but it is not practical when some vehicles are not connected. In a mixed platoon, the ability of CVs to measure the location of all vehicles depends on the MPR. At lower MPRs, all vehicles' locations are not available. 

Instead, a more feasible approach is to monitor the occupied cells by analyzing the cell density values at each time step to identify vehicles and predict their lane changes. By examining the occupied cells, it becomes possible to approximate vehicle locations. Additionally, considering that certain connected vehicles (CVs) measure speeds and distances, these values can aid in more accurately identifying vehicles. The process involves first using cell density monitoring to determine the presence of a vehicle within a cell. Second, if a vehicle is identified within that cell, and if measured speed and distance data are available within the cell, these measured values are utilized to enhance the accuracy of predictions. In cases with no measured values available, the cell densities and speeds are translated into the location and speed of that vehicle, which may cause some errors.

When targeting a specific vehicle, it becomes possible to monitor several cells ahead to identify the leader vehicle in an occupied cell. Utilizing Equation \eqref{eqrhomeasure} provides the estimation of the approximate location of the leader. Once the locations are determined, the cell speed is used to calculate the leader's speed and subsequently determine the safe speed from Equation \eqref{eqsafespeed}. Finally, the benefit value is calculated based on the safe speeds in the current lane and the target lane. In the case of no leader captured or gap amount is higher than the effective range, we assume that the preceding or leader vehicle is far enough to let the targeted vehicle drive without any impact of the preceding vehicle. Therefore, the leader speed is assumed to be the free flow speed. 

During the process, when the cumulative benefit memory exceeds the threshold, the lane change model indicates a probable lane change at that time step. For example, suppose that the actual benefit memory for a certain vehicle at $t=15$ s is 1. At the current time step, the lane change model computes the benefit values for the next 6 s with a time step of $dt = 0.5$ s. Assuming the list of computed benefits for each time step is [0, 0.1, 0.2, 0.2, 0.3, 0.25, 0.3, 0.2, 0.3, 0.3, 0.3, 0.3], and the threshold is set to 2. Therefore, the cumulative benefit for the given duration is [0, 0.1, 0.3, 0.5, 0.8, 1.05 , 1.35, 1.55, 1.85, 2.15, 2.45]. The cumulative benefit memory value exceeds 2 within the next 4.5 seconds. This means that the model predicts a lane change for this particular vehicle at $t=19.5$ s. However, it is important to note that not all vehicles may experience a lane change if their benefit memory does not exceed the threshold.

The final step of the LC model involves checking for sufficient space in the target lane. As the location of the lane changing vehicle (target vehicle) is estimated, the cell occupancy in the target lane is examined to ensure there is ample room for the vehicle to change lanes. To check that, the cell density is evaluated taking into account the cell length, minimum gap, and a minimum space required for the lane-change vehicle. A maximum density is determined for the mentioned parameters in the scenario. If there is insufficient space, the vehicle will need to wait until the next time step and postpone the lane change until enough space becomes available.

\textcolor{blue}{
\subsubsection*{Impact of the number of lanes on traffic prediction}
Lane change prediction and the entire control framework apply to roads with any number of lanes. Examples in this section and the results section describe a two-lane road section for simplicity. Since the model predicts in-flow and out-flow of lane changing for the target lane, a lane change from the target lane into adjacent lanes (the lanes on the right-hand and left-hand sides of the target lane) and vice versa can contribute to changes in the traffic states (refer to Equations \eqref{eqQtot} and \eqref{eq:dn/dt} where lane change flow is can take values depending on numbers of lanes). Thus, applying the proposed framework for all lanes of the road results in considering every lane of the road. Indeed, each individual lane is modeled to monitor the ingress and egress of vehicles independently by looking at the traffic states of itself and adjacent cells only.
}

\section{Optimal control} \label{optimalcontrolsec}
In this section, a control strategy is described with the primary objective of minimizing energy consumption while ensuring passenger comfort and providing the required mobility. To align with the traffic model, the speed is controlled over the optimal control horizon. Therefore, the optimization problem focuses on determining the optimal acceleration value, making it the decision variable for each time step. Since the optimization problem has to be solved for the prediction horizon, the decision variables includes a list of the acceleration values for the entire time steps during the prediction horizon. Moreover, speed and position for each time step are calculated from the acceleration using the vehicle dynamics.

This study uses the Sequential Least Squares Programming (SLSQP) method \cite{SLSQPbook} to solve the optimization. The objective function represents the required power to overcome resistance, provide acceleration, and grade \cite{parametric}. It consists of linear, quadratic, and cubic terms which results in a non-linear objective function. To handle the cubic term in the objective function, a linear approximation is employed, with minor error in the energy consumption equation. 

The optimal control strategy operates in coordination with the traffic flow model and the LC model, which are used in traffic constraints. CVs measure and transmit traffic information, which is then filtered and estimated using the UKF. The LC model, together with the modified PW model, aims to predict lane changes for each cell and subsequently forecast cell density and speed. This prediction process generates trajectories for preceding vehicles. By obtaining these trajectories, the control strategy establishes constraints and adjust the speed of the target CV in the upcoming seconds.

\subsection{Vehicle dynamics}
Since the control variable is acceleration, the vehicle's dynamics including the position and speed are calculated by kinematics:
\begin{equation}
    v(k+1) = v(k) + a(k) \cdot dt \\
    \label{eqdynamcisspd}
\end{equation}
\begin{equation}
\label{eqxstate}
    x(k+1) = x(k) + v(k) \cdot dt
\end{equation}
where $k$ is the time stamp and $x(k)$ is the longitudinal location of the target CAV at time $k$. \textcolor{blue}{ The original form of Equation \eqref{eqxstate} has an additional term of $\frac{1}{2} a(k) \cdot dt^2$ which is dropped in this study. Due to the short-time step assumption, this term is small even in high acceleration. Minor estimation errors are included in the following sections in car-following constraints.}

\subsection{Objective function}
The main objective of the control strategy is to minimize energy consumption during traveling, while also ensuring a comfortable driving experience. To achieve this, penalty terms are introduced for the acceleration value, and the rate of acceleration changes. \textcolor{blue}{To model the acceleration change into an optimization problem, large longitudinal jerk will be penalized to ensure that the vehicle can reasonably adjust the acceleration and maintain driver comfort}. The energy consumption rate is calculated using a well-known parametric road load equation, which incorporates the principles of vehicle dynamics based on fundamental physics laws \cite{parametric}. Vehicle powertrain principles are not in the scope of this study but drivability and vehicle capabilities are considered as constraints. Some power demands such as accessory power (e.g. driver comfort and air conditioning) are not mentioned in the objective function since they have a minor effect on the predicted driving strategy and travel time. Therefore, the objective function only consists of terms that have the most impact on the speed control. 
\begin{equation}
    \label{eqobjective}
     J = \min \int_{t= t_0}^{t_f} (P_{\mathrm{aero}} + P_{\mathrm{accel}} + P_{\mathrm{roll}} + P_{\mathrm{grade}} + w_1 a^2 + w_2 (\Delta a)^2 + w_3 \cdot s_1^2 +  w_4 \cdot s_2^2 ) dt \
\end{equation}
$P$ terms stand for the required power with respect to the time. $P_{\mathrm{aero}}$ is aerodynamic drag power (air resistance), $ P_{\mathrm{accel}}$ is acceleration power, $P_{\mathrm{roll}}$ is rolling resistance power, $P_{\mathrm{grade}}$ is road grade required power, and \textcolor{blue}{$ \left(\Delta a(k) \right)^2 = \left( a(k) - a(k-1) \right)^2$ is the longitudinal jerk between two consecutive time steps}. The objective function is discretized using the pseudo-spectral method \cite{pseudo-spectral} for each time step in the prediction horizon, considering that the trajectories are available at discrete times.
\begin{equation}
\begin{aligned}    
    \label{eqobjective2}
     J = \min \sum_{k=0}^{N_H-1} \left[ 0.5\rho C_D a(k) v(k)^3 + C_{RR}m_{t} g v(k) + k_m m_{t} a(k) v(k) + m_{t}gZv(k) \right.\\ 
     \left. + w_1 a(k)^2 + w_2 \left(\Delta a(k) \right)^2 + w_3 \cdot s_1(k)^2 +  w_4 \cdot s_2(k)^2 \right]
\end{aligned}
\end{equation}
where $N_H$ is the number of steps in prediction horizon, $v(k)$ is the vehicle speed at time step $k$ (m/s), $a(k)$ is the vehicle acceleration ($m/s2$), $\rho$ is the density of air ($\sim   
  1.2kg/m3$), $C_D$ is the aerodynamic drag
coefficient, $C_{RR}$ is the rolling resistance coefficient, $m_{t}$ is the
total vehicle mass ($kg$), $g$ is the gravitational acceleration ($9.81m/s2$), $Z$ is the road gradient
($\%$) and $k_m$ is a factor to account for the rotational inertia of the powertrain. We assumed a value of $k_m$ = 1.1 \cite{plotkin}. $w_1$, $w_2$, $w_3$, and $w_4$ are positive constant weights for the penalty term. $s_1(k)$ and $s_2(k)$ are slack variables to be penalized and are presented in the constraints section. At each time step, the energy consumption rate is calculated starting from actual time and up to the prediction horizon (10s) which results in the vehicle power. The lateral movement is treated as normal driving behavior and only longitudinal acceleration is a metric of energy efficiency. 

\subsection{Constraints}
A target vehicle is subject to two types of constraints: physical boundaries and traffic constraints. Both types of constraints impose limits on the vehicle's speed and acceleration, ensuring safe and compliant operation within the given traffic environment.

(i) Physical bounds
The physical boundaries take into account the vehicle's dynamics and capabilities, ensuring that it operates within its physical bounds such as speed and acceleration.
\begin{equation}
    \label{eqspdbound}
    v_{\min} \le v(t) \le v_{\max}
\end{equation}
\begin{equation}
    \label{eqaccbound}
    a_{\min} \le a(t) \le a_{\max}
\end{equation}
where $v_{\min}$ is set to be 0, $a_{\min}$ represents the maximum deceleration rate, and $a_{\max}$ is the maximum acceleration.

(ii) Traffic constraints
 The traffic constraints govern the vehicle's behavior in accordance with traffic principles, including maintaining a reasonable car-following distance and adhering to traffic signal rules. First, the target vehicle must maintain a safe distance with the preceding vehicle to avoid collision, in addition to a maximum distance to ensure mobility \cite{energyandmobility}. Therefore, given the availability of a preceding vehicle, we enforce the constraints:
 \begin{equation}
     d(t) \ge d_p(t) + \beta \cdot \sigma[d_p(t)] - d_{\max}  - s_1(t)
    \label{eqdmax}
 \end{equation}
\begin{equation}
    \label{eqdmin}
     d(t) \le d_p(t) - \beta \cdot \sigma[d_p(t)] - (d_{\min} + h_{\min} \cdot v(t) ) + s_2(t)
\end{equation}
\begin{equation}
    s_1(t) , s_2(t) \ge 0
\end{equation}
where $d(t)$ is the target vehicle location at time t and $d_p(t)$ is the preceding vehicle location which is available from the predicted trajectory. In other words, $d_p(t) - d(t)$ is the distance between the target vehicle and the preceding vehicle. $\beta$ is the confidence level which is assumed to be 1 according to \cite{ShaoandSunb}. $d_{\max}$ is the maximum allowable distance to the preceding vehicle to ensure mobility and the traffic throughput, $d_{\min}$ is the minimum distance between vehicles to avoid collision (bumper to bumper), $h_{\min}$ is the minimum headway in seconds, and $ \sigma[d_p(t)]$ is the standard deviation of the estimated location of the immediate preceding vehicle. For computation methods and further explanations, readers are referred to \cite{ShaoandSunb, ecoapproach}. Finally, $s_1(t)$ and $s_2(t)$ are positive slack variables that help control and avoid invisibility when values are close to the equation boundaries and are penalized in the objective function.

When the target vehicle is within the range of V2I communications, additional information becomes available, enhancing energy efficiency due to the presence of SPaT messages. Assuming that the location and speed of the target vehicle, and the distance left to the signal are available, \textcolor{blue}{the control strategy aims to ensure that the target CAV arrives at the intersection within a critical time, i.e. the latest time that target CAV can arrive to the signal location based on the signal phasing plan. This critical time, combined with the distance left to the intersection, determines the target CAV's lower bound speed referred to as $v_{\mathrm{optimal}}$}. The approach addresses two main requirements. First, it ensures mobility when there is no preceding vehicle, and as a result, there is no maximum spacing constraint to implement the Equation \eqref{eqdmax}. Second, considering the signal phases, the target CAV can adjust its speed to reach the intersection at the nearest green light. Since the decision variable is the acceleration, and the speed can be directly computed by \eqref{eqdynamcisspd}, the speed control is constrained by an optimal speed:
\begin{equation}
    \label{eqspdspat}
    v_{\mathrm{optimal}} \le v(t) 
\end{equation}
where $v_{\mathrm{optimal}}$ is an optimal speed that is computed from the described critical time and distance remaining to the signal. Employing this lower bound for the speed guarantees that the vehicle will generate a plan to arrive at the closest possible green light. Simultaneously, maximum and minimum spacing constraints in Equations \eqref{eqdmax} and \eqref{eqdmin} ensure the reliability of the model in a fast-changing flow case.

Furthermore, when the signal is in the green phase and the target CAV is in the last cell left to the signal, the optimal speed is designed to uniformly increase to reach the free flow speed at a predetermined rate which is based on the free flow speed value. The main reason is to ensure the vehicle will pass green light with no latency. In this case, the value of $v_{\mathrm{optimal}}$ is small due to small distance left while a longer green time is available particularly if the vehicle arrives at the beginning of the green phase. Therefore, if the car-following spacing to the preceding vehicle is smaller than $d_{\max}$ from Equation \eqref{eqdmax}, the framework decides to slow down until last seconds of green light. However, when the target CAV is in other cells other than last cell, the constraints in Equations \eqref{eqdmax} and \eqref{eqspdspat} balance the car-following spacing along with passing the nearest green phase.

The target vehicle can only pass through the intersection when the signal is in the green phase \cite{energyandmobility}. The predicted location of target vehicle at the time when signal is red, should be after the signal location. Additionally, in the traffic flow model, it is assumed that the speed of the last cell is equal to zero when the signal is in the red phase.
\begin{equation}
    \label{eqdsig}
    d(t_{\mathrm{green}}) \le d_{\mathrm{signal}} \le d(t_{\mathrm{red}}) 
\end{equation}
$t_{\mathrm{green}}$ and $t_{\mathrm{red}}$ represent the starting times for the next green and red lights, when the eco vehicle is close to the intersection. $d_{\mathrm{signal}}$ is the location of the signal, $d(t_{\mathrm{green}})$ and $d(t_{\mathrm{red}})$ are the predicted locations of the target CAVs at that certain time stamp.

As mentioned, the optimal control strategy can take into account a connected signalized intersection. When an target vehicle is not within range of SPaT messages, it relies on Equations \eqref{eqspdbound}--\eqref{eqdmin} for its decision-making. However, when the eco vehicle can receive real-time signal timing information through the V2I connection, it can adjust its speed to find an optimal speed for traversing the intersection. For instance, if the eco vehicle determines that it can only reach the current green phase by driving at maximum speed, it plans the speed control accordingly. On the other hand, if the current signal phase is red, the eco vehicle aims to slow down in order to delay its arrival time at the intersection, allowing it to potentially catch the next green phase. To incorporate SPaT messages into the optimal control when the connected intersection is within range, an optimal speed is computed based on the remaining distance to the intersection. This ensures the eco vehicle's best performance in passing the intersection. Equation \eqref{eqspdspat} is introduced to determine the minimum speed required to ensure that the eco vehicle reaches the earliest feasible green phase, replacing the maximum car-following distance in Equation \eqref{eqdmax}. This modification allows the eco vehicle to maintain higher distances while guaranteeing mobility and traffic throughput. In summary Equation \eqref{eqcons} specifies the constraints for this problem:
\begin{equation}
    \label{eqobjective2}
\begin{aligned}    
     J = \min \sum_{k=0}^{ \color{blue} N_H-1} \left[ 0.5\rho C_D a(k) v(k)^3 + C_{RR}m_{k} g v(k) + k_m m_{t} a(k) v(k) + m_{t}gZv(k) \right.\\ 
     \left. + w_1 a(k)^2 + w_2 \left(\Delta a(k)\right)^2 + w_3 \cdot s_1(k)^2 +  w_4 \cdot s_2(k)^2 \right]
\end{aligned}
\end{equation}

s.t.
\begin{equation}   
    \begin{cases}
        \label{eqcons}
        \text{Equations \eqref{eqspdbound}, \eqref{eqaccbound}, \eqref{eqdmax}, \eqref{eqdmin}}  &  \lvert d_{\mathrm{signal}} - d(t_0) \lvert > \text{SPaT range}  \\
        \text{Equations \eqref{eqspdbound}, \eqref{eqaccbound}, \eqref{eqdmin}}, \eqref{eqspdspat} , \eqref{eqdsig}   &  \lvert  d_{\mathrm{signal}} - d(t_0) \lvert \le  \text{SPaT range}    \\
    \end{cases}
\end{equation}
where $d(t_0)$ is the location of the target CAV at the start time of optimization that is available for a given CAV.

\section{Numerical results} \label{resultssec}
This section presents the results of implementing the proposed framework on scenarios. The entire framework is implemented in the simulation environment to study the energy efficiency performance. Increasing network complexity or various situations in generated scenarios has minor effects on the framework and the traffic prediction itself. Therefore, scenarios consist of a simplified network in order to highlight the impact of LCs and provide an example of the proposed method. 

The numerical results section begins by presenting the created scenarios and assumptions employed in the simulation. Subsequently, speed control is evaluated before employing LC prediction. The performance of the entire platform (speed control + LC prediction) is then presented, specifically in terms of energy efficiency, highlighting the benefits achieved, which serve as the main motivation for this study. It includes evaluation of parameters in the LC prediction model, assessment of various traffic volumes, and various MPRs energy benefits. Furthermore, in order to account for potential side effects of the control strategy, the travel time and trajectory of the targeted CAVs are assessed. Finally, the precision of the traffic prediction component is studied as a secondary concern.

\subsection{Generated scenarios, parameters and assumptions}
The presented method is implemented in a traffic simulation to calculate the energy consumption and compare it to the models that do not consider the impact of lane changing. SUMO (Simulation of Urban MObility) software is utilized for the implementation of the mechanism. To retrieve values, program, and control the target vehicles, the ``Traffic Control Interface'' (TraCI) is used, which provides online access to a running simulation \cite{sumo-traCI, sumo-traCI2}. TraCI is a client-server interface that is used via Python programming language in the current study. For more detailed information, readers are referred to \cite{sumo-traCI}. In summary, a network is created along with a defined traffic demand. The traffic demand enters the network and is updated at every time step. Information regarding the vehicles, lanes, signals, and other elements can be retrieved from the simulation at each time step. Furthermore, values such as dynamics and characteristics for the elements can be programmed and changed dynamically for each time step. Fig. \ref{picsumo} represents a section of a scenario including 150m before and after the signal. The red vehicle is the target CAV that is controlled by Equation sets \eqref{eqobjective2} and \eqref{eqcons}. As depicted, since the traffic light is in red phase and cells ahead have lower amounts of speed, the controlled vehicle slows down earlier and as a result, maintains a larger gap. The larger gap induces a lane-change which is mentioned by an arrow in the figure. The simulation procedure is explained as follows:

\begin{figure}
    \centering
    \includegraphics[scale=0.36]{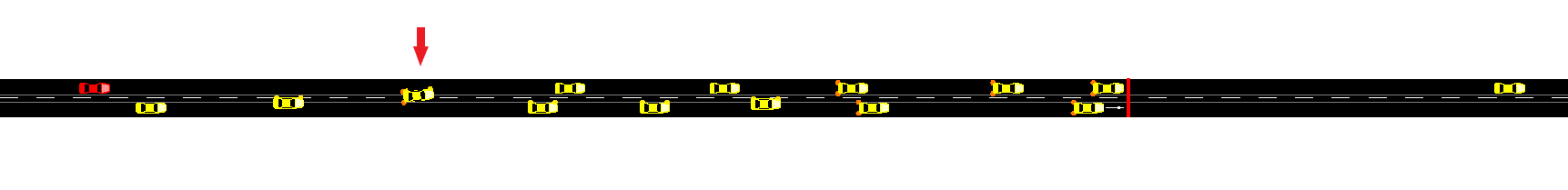}
    \caption{A section of the scenario, modeled traffic stream are loaded on the network, and red vehicle is the target CAV controlled by the optimal strategy}
    \label{picsumo}
\end{figure}

\subsubsection*{Network}
The network consists of a two-lane, one-way road with a signalized intersection. To focus on the impact of lane changing, the signal has only two phases: one 20 s red phase and one 20 s green phase. The yellow phase is assumed to be 2 s and is modeled in the network. But, it is considered as part of the red phase for the traffic prediction modeling. The road section can either be a long road segment or include a warm-up section. However, the most significant section is within the range of the connected intersection where signal timing can be received. The V2I range is assumed to be 350 meters. The network includes a 150 m section before the range, a 350 m section within the range and preceding the signal, and a 350 m section after the signal and within the range. The right lane is labeled as '1', and the left lane is labeled as '2'. The maximum speed for both lanes is 15 m/s for this network.

Traffic demand is defined and applied to the generated network. To model lane changing due to speed gain, the right lane (lane 1) is assumed to be the slower lane, while the left lane (lane 2) serves as the fast or target lane for lane changing vehicles. In lane 1, a slow vehicle is introduced into the scenario frequently to create sufficient motivation for lane changing, simulating behaviors such as buses or recreational cruising vehicles. Each element in the traffic flow within the SUMO simulation must be assigned one of the predefined vehicle types. \textcolor{blue}{ For this study, three vehicle types are defined: 1) regular HVs and slow vehicles which are the same type with only slower maximum speed for slow vehicles, 2) CVs, and 3) the target CAV. HVs, slow vehicles, and CVs are driven by humans while the CAV is controlled by the optimal control autonomously. Table \ref{tabledemand} provides a further explanation of how vehicles are categorized in each lane.}

\subsubsection*{Demand}
To ensure a realistic traffic simulation, vehicles are introduced into the scenario randomly based on their departure time. For each MPR and traffic volume, 20 various flow set plans are randomly generated rather than having uniform traffic. The plan involves initiating a specific number of vehicles within a defined time interval, with the departure time of each vehicle being randomized in this process. Therefore the time headway and spacing between each two consecutive vehicles is a random number. Seeds are also employed to facilitate scenario reproducibility. The introduction of randomness in vehicle departures results in diverse vehicle spacing and a range of lane-change phenomena, enabling a robust evaluation of the LC model.

To investigate the influence of congestion and traffic volume on the method performance, 6 various traffic flow sets, ranging from low volume to high volume, are provided for both lanes. Based on the scenario, a certain number of vehicles depart over 50 seconds which is subsequently translated to the traffic volume.

\begin{table}[pos=htp!]
\begin{center}
\caption{\textcolor{blue}{Vehicle types modeled in two lanes}}
\label{tabledemand}
\begin{tabular}{ | m{5em} | m{1cm}| m{1cm} | m{1cm} | m{1cm} |} 
  \hline
  Lane/Type & CAV & CVs & HVs & Slow HVs \\ 
  \hline
  Lane 1 &  --- & \checkmark & \checkmark & \checkmark \\ 
  \hline
  Lane 2 & \checkmark & \checkmark & \checkmark &  --- \\ 
  \hline
\end{tabular}
\end{center}
\end{table}

The traffic volume in each lane is composed of vehicles of different types (Table \ref{tabledemand}), and the ratio of CVs to all vehicles determines the MPR in each lane. When defining the traffic demand for the lanes, efforts are made to maintain a similar MPR for both lanes. This approach ensures a balanced representation of CVs in the traffic flow and enables a fair comparison.

\subsubsection*{\textcolor{blue}{Parameters in traffic models, optimization, and simulation}}

\textcolor{blue}{
There are various parameters in the presented equations and simulation setup that are defined in Table \ref{tableparams}.}

\begin{table}[pos=htp!]

\begin{center}
\caption{Traffic modeling, optimization, and simulation parameters simulated in results}
\label{tableparams}
\begin{tabular}{ | m{1.5cm} | m{1.5cm}| m{1cm} | m{1cm} | m{1cm} | m{1.5cm} |  m{1.5cm} |} 
  \hline
  $v_0 (m/s)$ & $a_{max} (m/s^2)$ & $b(m/s^2)$ &  $\tau (s)$ & c & $gap_{min} (m)$  & $\rho_{jam} (v/km)$\\ 
  \hline
  $15$ & $2.6$ & $4.5$ & $1$ & $10.14$  & $2.5$ & $130$ \\ 
  \hline

  $h_{min} (s)$  & $d_{min} (m)$ & $d_{max} (m)$ & $m_t (kg)$ & $C_{RR}$ & $w_1$ & $w_2$\\ 
  \hline
  $1.5$ & $2.5$ & $5 v_0$  & $1550$ & $0.8$ & $3000$ & $150$ \\

  \hline

\end{tabular}
\end{center}
\end{table}

\textcolor{blue}{
Weights for slack variables are limited to 5\% of weights on decision variables. Moreover, since traffic state prediction employs UKF, the range of traffic state variables is described in \cite{wang2005real}. Furthermore, driver behavior and vehicle variables that are not described in Table \ref{tableparams}, are default values of the SUMO simulator \cite{sumo-traCI2}.}

\subsubsection*{Method implementation process}

The framework is implemented in a discrete space and time environment which divides each segment into smaller cells. Cell length ($\Delta x$) is chosen to be 15 meters and the time step ($\Delta t$) is 0.5 seconds. The reason for choosing the values for cell length and time step is to have a balance between prediction accuracy and computation time. Shorter intervals will improve the prediction accuracy, particularly in higher MPRs but increase the processing time. The average processing time is 0.14 seconds per time step ($\Delta t$ = 0.5 s) for the following scenarios.

The modified PW model predicts cell states for the next 10 seconds, while the LC model forecasts LC occurrences only 4--6 seconds in advance. It has been observed that lower prediction horizons yield reduced energy benefits, while higher horizons result in inaccuracies and overestimation in LC predictions. The optimal control computes acceleration values for the same 10-second horizon as the PW model, but with time steps of 0.5 seconds, compared to the traffic model's 1-second time steps. Among the computed accelerations for the next 20 time steps, only the acceleration of the first and second time steps will be used for the control. Subsequently, the optimization process is rerun to obtain updated optimal values. This iterative process ensures efficient speed adjustments, and fuel consumption is computed accordingly for each time step. The traffic prediction model and also the optimal control are updated every 1 second.

The target lane for lane changes which is specifically lane 2 in scenarios, is intentionally designed to be less congested in the generated scenarios. This configuration ensures the availability of sufficient gaps and opportunities for successful lane changes. The instantaneous nature of lane changing in the SUMO simulation makes capturing fractional vehicle locations not possible. Meaning that the vehicle in each time step only belongs to a certain lane not proportion of it in a lane. This is compatible with the stated convention in Section \ref{LCdensitysec}.

Moreover, the algorithm takes into account the influence of the red light in Equation \eqref{eqVe} on the last 6 cells, equivalent to a distance of 90 meters.

In the following sections, each data point represents a mean of 20 randomly generated scenarios. We report the average effect on the target CAVs and exclude the effects on other vehicles. Furthermore, the error bars are the standard deviation of these 20 scenarios. Moreover, the energy consumption of the optimal speed control generated by the standard PW model without LC prediction represents the baseline.

We have observed that the positioning and sequence of the target CAV within the platoon exert a notable influence on the energy advantages achieved. Factors encompassing the gap, preceding vehicle characteristics, and the likelihood of lane changes ahead of the target CAV significantly contribute to these energy benefits. To consider various situations and cover the impact of various lane changes in the platoon, the control is applied to the last CAV in the formation. Adding multiple target CAVs was also an available option but the speed control of a vehicle will affect all following vehicles. Therefore, in addition to the LC impact, latter target CAVs are also influenced by the earlier target CAVs which is not the main goal of the study. 

\subsection{ \textcolor{blue}{Energy benefit of speed control with no LC prediction}}
\textcolor{blue}{
One key element of the proposed optimal control is the speed control as described in Equations \eqref{eqobjective2} and \eqref{eqcons}. This optimization problem is integrated with traffic prediction and LC prediction to develop a control framework. The first step of evaluation is to validate the speed control itself. Therefore, we only employ traffic state prediction without LC prediction to provide the trajectory of the preceding vehicle. This trajectory will be used for the optimization problem. This case generally evaluates the performance of the baseline.}

\begin{figure}
    \centering
        \includegraphics[width=0.8\textwidth]{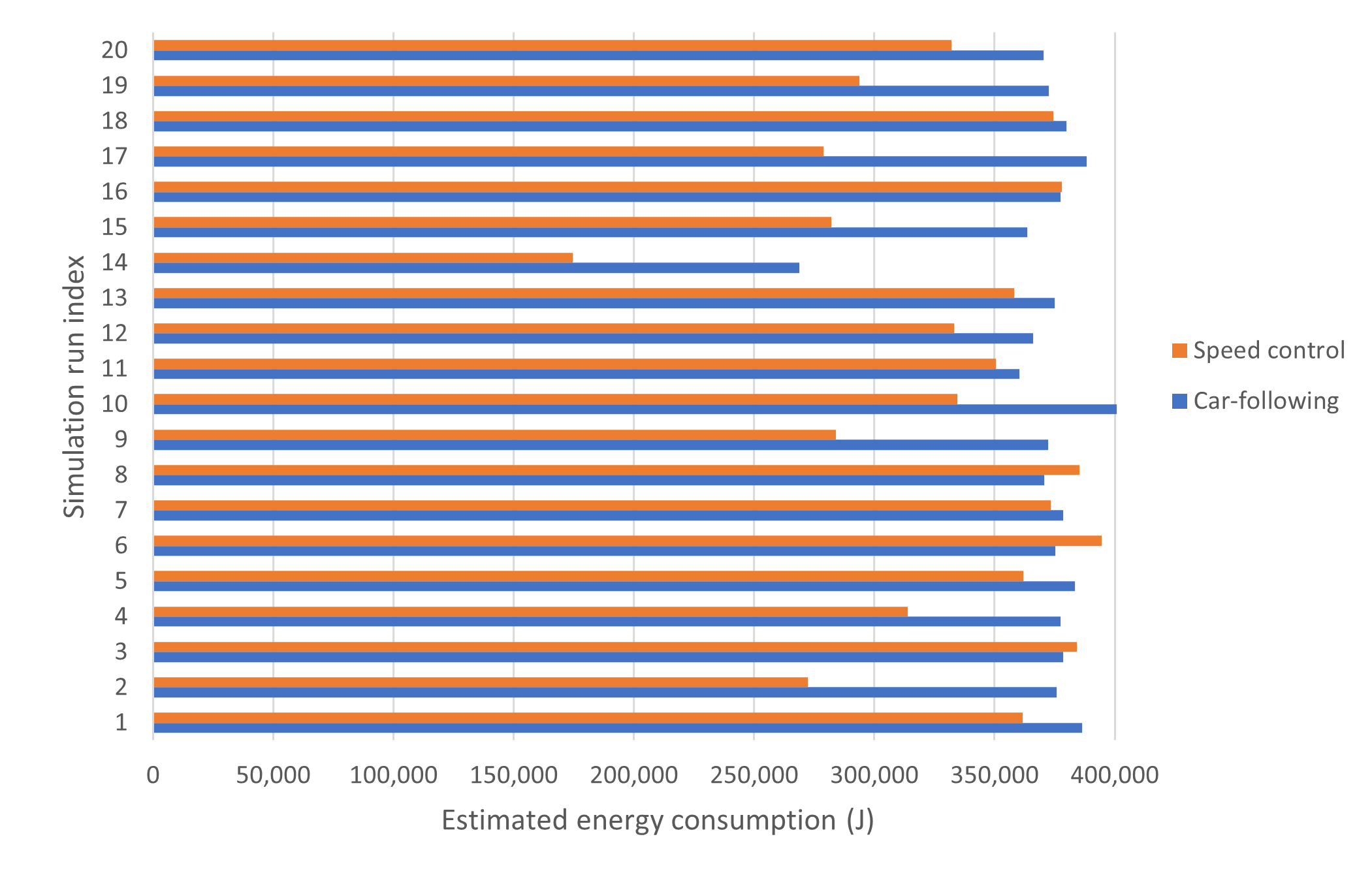}
    \caption{Target CAV energy consumptions at 70\% MPR for the speed control compared to Krauss car-following model for 20 generated scenarios}
    \label{picspeedcontrol}
\end{figure}

\textcolor{blue}{
Figure \ref{picspeedcontrol} depicts the energy consumption of two models. The orange chart represents speed control with no LC prediction and the blue chart shows energy consumption for the Krauss car-following model which is the default car-following model of SUMO \cite{krauss1998microscopic}. Consumption values are shown for 20 randomly generated scenarios with various departure time for all vehicles. On average, speed control consumes 10.6\% less energy. The following sections include LC prediction and focus on the impacts of speed control together with LC prediction.}

\subsection{LC model performance}
In this section, we evaluate the performance of the LC prediction model in Equation \ref{eqbenefit} when coordinated with the modified traffic flow model in Equations \eqref{eqmodifyv} and \eqref{eqmodifyrho} and find the LC model parameter. The threshold value in the LC model plays a crucial role in determining LC predictions. Lower threshold values indicate a more sensitive model that predicts a higher number of lane changes and LC density overestimation from Equation \eqref{rhoLCestimation}. On the other hand, larger threshold represent a less sensitive model which can result in two phenomena at high values: first, zero LC prediction and second, delayed LC prediction which has a negative impact.

We analyze the energy consumption of the system based on different threshold values to figure out the appropriate values for the model. Thus, we select various thresholds with increments of 0.5 and record the corresponding energy benefits. Where the benefit values are close and relatively large, threshold increments are 0.25 to identify the peak more accurately. The traffic volume is the flow set which has the highest energy benefit that presented in the following subsection.

\begin{figure}
    \centering
        \includegraphics[width=0.9\textwidth]{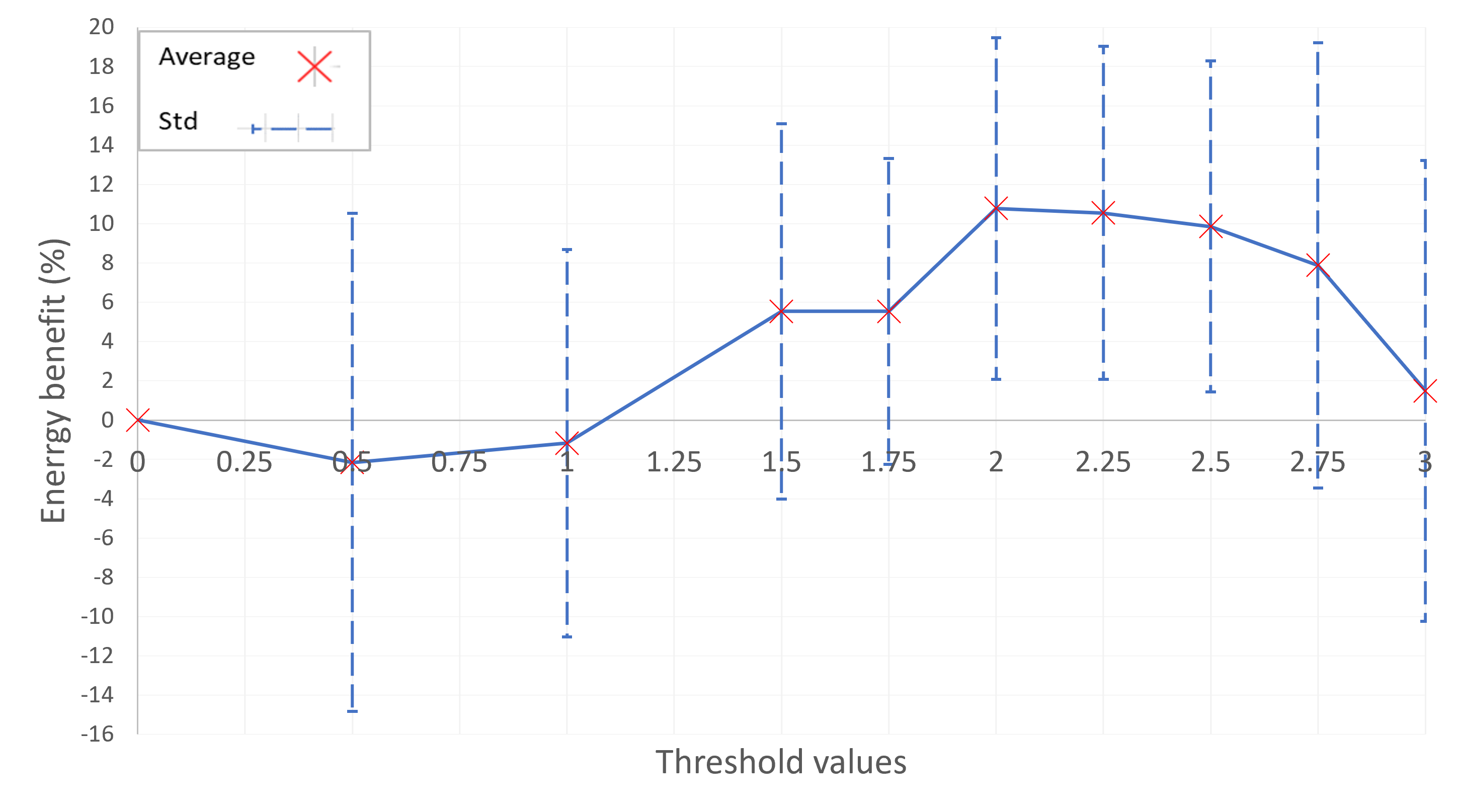}
    \caption{Target CAV energy benefits at various threshold rates of the LC prediction model}
    \label{picthreshold}
\end{figure}

Figure \ref{picthreshold} demonstrates the energy benefit of the modified framework compared to the baseline at 70\% MPR. As depicted, values between 2 and 2.5 result in more energy efficiency which is due to better LC prediction. False LC predictions misguides the optimal control and results in less energy benefit and even energy loss compared to the baseline. It is observed that for threshold values of 0.5 and 1, the modified platform could result in disadvantage. Due to a higher proportion of incorrect predictions, there is greater uncertainty and standard deviation in these threshold values as well. Moderate threshold values between 1.5 and 2.5 result in relatively lower fluctuations, indicating a more robust LC model. On the other hand, values of 2.75 and higher tend to underestimate the occurrence of lane-change events, leading to reduced energy benefits as some LC phenomena are missed or delayed in the prediction. Additionally, it is important to note that a threshold value of 0 is not practically applicable and is assumed to represent the baseline scenario where the LC model is not activated.

\subsection{Energy benefits for various traffic volumes}
In order to encompass a wide range of traffic volumes, a comprehensive set of flow sets are defined, ranging from 800 to 2000 vehicles per hour per lane. The created traffic flow sets represent various traffic volumes in the scenarios. For smaller volumes, larger gaps are observed, while congested sections are observed in larger volumes. When the traffic volume is small, there are scenarios with no LCs since vehicles can drive at free-flow speed and leave the scenario before reaching a slow vehicle, providing no reason to change lanes for speed gain. Even if there is a slow vehicle nearby, there are fewer vehicles that perform LCs in small volumes. Moreover, gaps are large enough to allow vehicles to change lanes freely with little impact on other vehicles. Consequently, predicting LCs has a minor impact on other vehicles in lower traffic volumes.

On the other hand, in higher volumes, the number of LC events increases due to the presence of more vehicles, while gaps become smaller. This results in the target CAV experiencing more frequent LCs immediately at the front and at closer distances. As a result, the LC prediction and control platform yield more significant advantages in higher traffic volumes. However at even higher volumes, there is insufficient gap in the target lane to execute LC. Consequently, in high volumes where LCs cannot occur due to shorter gaps, predicting LC will not yield much benefits.

\begin{figure}
    \centering
    \includegraphics[width=0.9\textwidth]{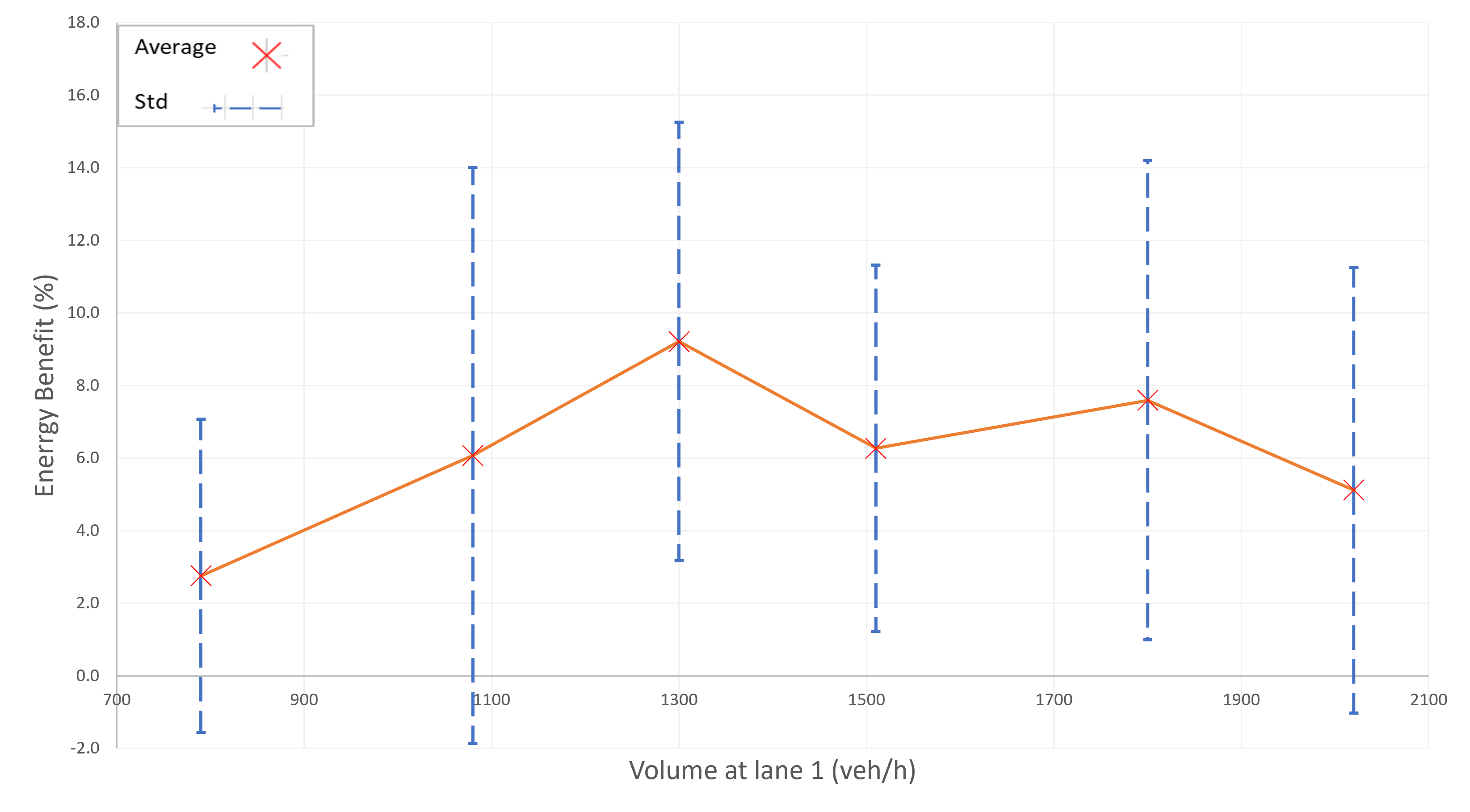}
    \caption{Target CAV energy benefit for various traffic volumes}
    \label{picvarious_flows}
\end{figure}

The traffic volumes significantly influence the frequency and impact of lane changes on the target CAV, which, in turn, affects the benefits of the LC prediction and control platform. Figure \ref{picvarious_flows} depicts the benefits of the presented platform compared to the baseline at the threshold value of 2.5 and 70\% MPR. As depicted, traffic volume of 1300 vehicles per hour which is a moderate volume has the highest energy benefit with average benefit of 9.2 \%. Either lower and higher volumes result in less energy efficiency due to the mentioned rationals.

\textcolor{blue}{The inconsistency in the descending trend from volume 1300 to 2000 vehicles per hour refers to a higher speed difference between two lanes in 1500 volume compared to volume 1800 and 2000 vehicles per hour. As LC prediction heavily relies on the speed difference between two adjacent lanes, it causes an overestimation in the number of LCs by the target CAV. Therefore, the vehicle slows down but since no vehicle changes the lane, it accelerates again. These cycles cause energy loss which reduces the benefit slightly. Although LC prediction based on the speed and density of the source and target lanes is extensively accepted and applied in studies \cite{LCmodel, knoop2010lane, knoop2012quantifying}, it might have such limitations. However, the model still shows improvement over the baseline.}

In traffic volumes below 800 vehicles per hour, the frequency of lane changes is naturally reduced due to less speed gain motivation as drivers in lane 1 and lane 2 have approximate safe speeds from Equation \eqref{eqsafespeed}. Similarly, for traffic volumes exceeding 2000 vehicles per hour, the frequency of lane changes is also diminished due to fewer available gaps and opportunities for lane changes in highly congested conditions.

\begin{table}[pos=htp!]
    \centering
    \caption{Dynamic and driver parameters for the defined vehicle types in simulation}
    \label{tableparameters}
    \begin{tabular}{ | m{2em} |c | c | c | c | c |c |} 
      \hline
      
      Type  & length($m$) & min gap($m$) & max deceleration($m/s^2$) & min time headway($s$) & imperfection\\ 
      \hline
      HV & 5 & 2.5  & 4.5 & 1 & 0.5 \\ 
      
      CV & 5 & 2.5  & 4.5 & 0.9 & 1 \\ 
      
      bus & 5 & 2.5 & 3 & 1 & 0.5 \\ 
      \hline
      \multicolumn{6}{|c|}{Maximum acceleration for all types is 2.6 $m/s^2$} \\
      \hline 
    \end{tabular}
\end{table}

\subsection{Energy benefit for various MPRs}
More CVs provide a greater amount of measurement and traffic information, leading to better cell density and speed estimation. As a result, more accurate safe speeds and gaps are computed using the Equation \eqref{eqsafespeed}, which in turn results in more precise LC predictions. This is expected to lead to higher levels of energy benefits.

In the simulations, both lanes are defined with similar MPR, and both CVs and HVs are allowed to change lanes. However, according to \cite{karbasi2023comparison}, CVs are defined slightly differently from HVs in terms of their driving behavior. This difference in driving behavior can cause CVs to make different lane-change actions compared to HVs, but both types of vehicles still perform lane changes due to speed gain. Table \ref{tableparameters} explains the distinctions in the definitions of the vehicles. The main differences in driving behavior include the time headway, which represents the driver's desired (minimum) time gap between vehicles, and the driving perfection, which indicates how skilled the driver is. For CVs, it is assumed that they are partially automated, leading to a lower time headway and a lower level of imperfection in their driving behavior.

\begin{figure}
    \centering
    \includegraphics[width=\textwidth]{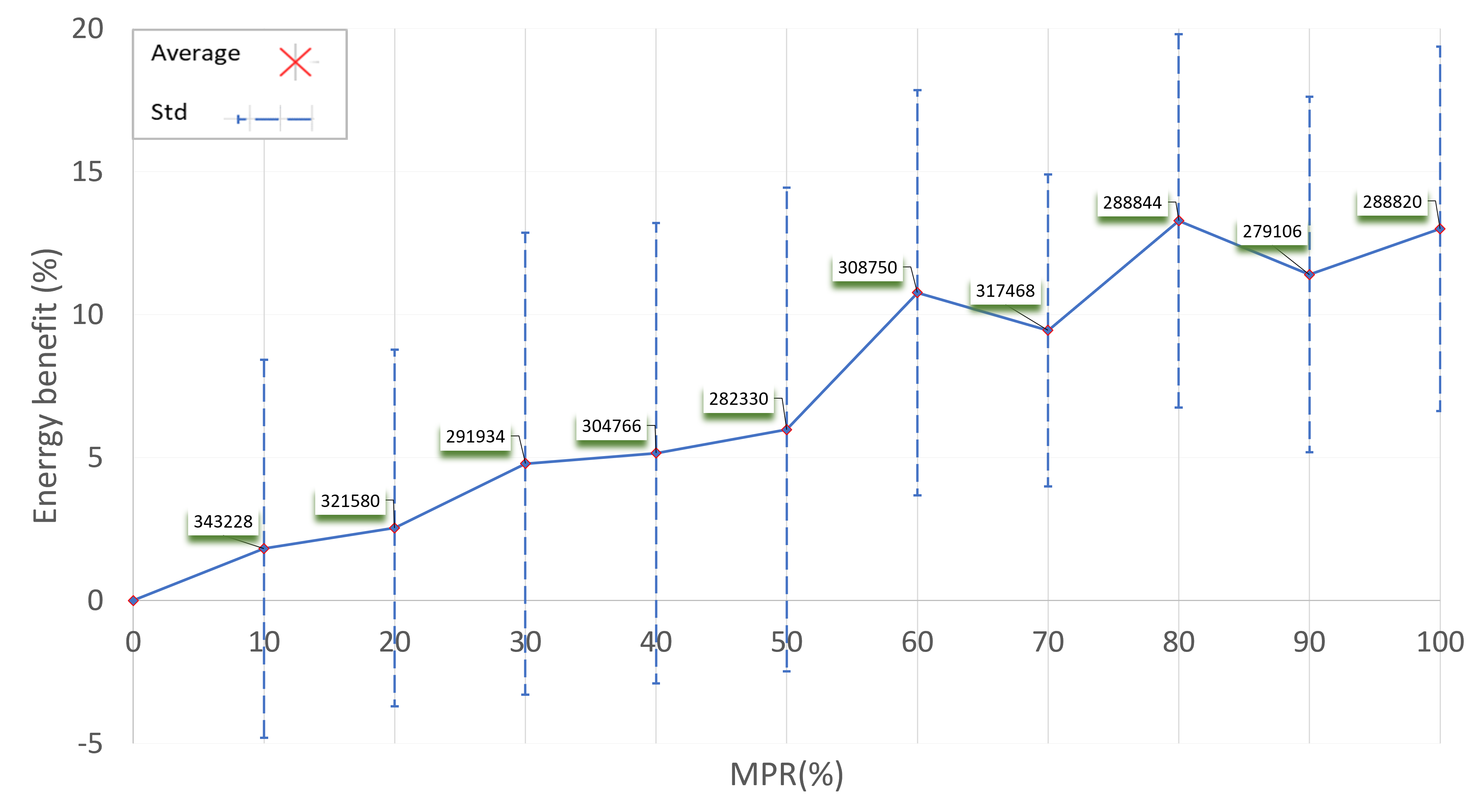}
    \caption{Target CAV energy benefit at various MPRs. Numbers on the plot represent energy consumption (unit is Jules) while employing speed control and LC prediction for the scenario}
    \label{picMPR_benefit}
\end{figure}

As previously mentioned, different traffic volumes result in varying energy benefits. To focus only on the impact of MPR, the traffic flow set with the highest energy benefit is selected for this part. Figure \ref{picMPR_benefit} demonstrates the energy benefit values for various MPRs with a model threshold of 2.5. Starting from low MPRs, there is no energy benefit due to the lack of data in the estimated cell states. The LC prediction model heavily relies on the cell density and speed estimations to compute the safe speed by Equation \eqref{eqsafespeed}. The cell density and speed also have a relation to the measured values which explained by Equations \eqref{eqrhomeasure} and \eqref{eqvmeasure}.
When estimations are not precise, the safe speeds in the two adjacent lanes become less accurate, and the benefit function from Equation \eqref{eqbenefit} does not represent actual LC occurrences. This scenario aligns with the baseline and results in minor energy benefit for MPRs close to 10\%. It is observed that for some random scenarios, even energy consumption increases slightly. This issue in the LC prediction can be addressed by filtering the LC predictions conservatively which is not in the scope of this study.

As the MPR increases, the LC prediction is based on more information provided by CVs, leading to higher energy benefits. The highest improvement compared to the baseline is an average of 13\%. Moreover, higher MPRs result in fluctuation reduction due to awareness of vehicles and their leader location.

Even in a partially connected platoon, the distances to non-connected vehicles and speeds can be measured. It is observed that after 60\% MPR, there is sufficient information about the vehicles states for the LC model which results in more than 10\% energy benefits for all MPRs over 60\%. 

As the MPR changes, the total number of available CAVs in the scenario also varies. For instance, the order and position of the target CAV at 10\% MPR differ entirely from those at 90\% MPR. To minimize fluctuations in the change of the target CAV, a strategy is employed where, for each pair of consecutive MPRs, the same CV becomes the target. For instance, in the generated scenario, the 7th CV in the platoon becomes the target CAV for both 90\% and 100\% MPRs. The reduction in benefits as the MPR increases can be attributed to this implementation.

\textcolor{blue}{Numbers in Figure \ref{picMPR_benefit} are the estimated energy consumption corresponding to energy benefit for 700 meters of the scenario (350 m before the signal and 350 m after). The values are estimated when employing LC prediction along with optimal speed control. EV's approximate energy consumption is about 150 Wh/km (540,000 Jules per kilometer) which is reasonable compared to our estimation \cite{ev-database}. Estimated energy consumption values are slightly lower than the actual energy consumed by a vehicle since auxiliary consumption (e.g. air conditioning, radio) is not considered in the presented estimation. Moreover, it is not expected to have a lower energy consumption for higher MPRs since the order and the position of the target CAV in the platoon change as MPR varies. In fact, we should compare energy consumption for each MPR independently with its baseline. Thus, the energy benefit of LC prediction working together with speed control should only be compared with the energy consumption of speed control without LC prediction (baseline) in each MPR.}

\subsection{Travel time assessment}
As discussed in the optimal control section particularly in the constraints (Equations \eqref{eqdmax} and \eqref{eqspdspat}), ensuring mobility is a critical aspect for any viable model. An optimal control framework that reduces energy consumption but leads to longer travel times for vehicles would be impractical as it could negatively impact the road capacity and user satisfaction. 

\begin{figure}[pos=htp!]
    \centering
    \centering
    \includegraphics[width=0.9\textwidth]{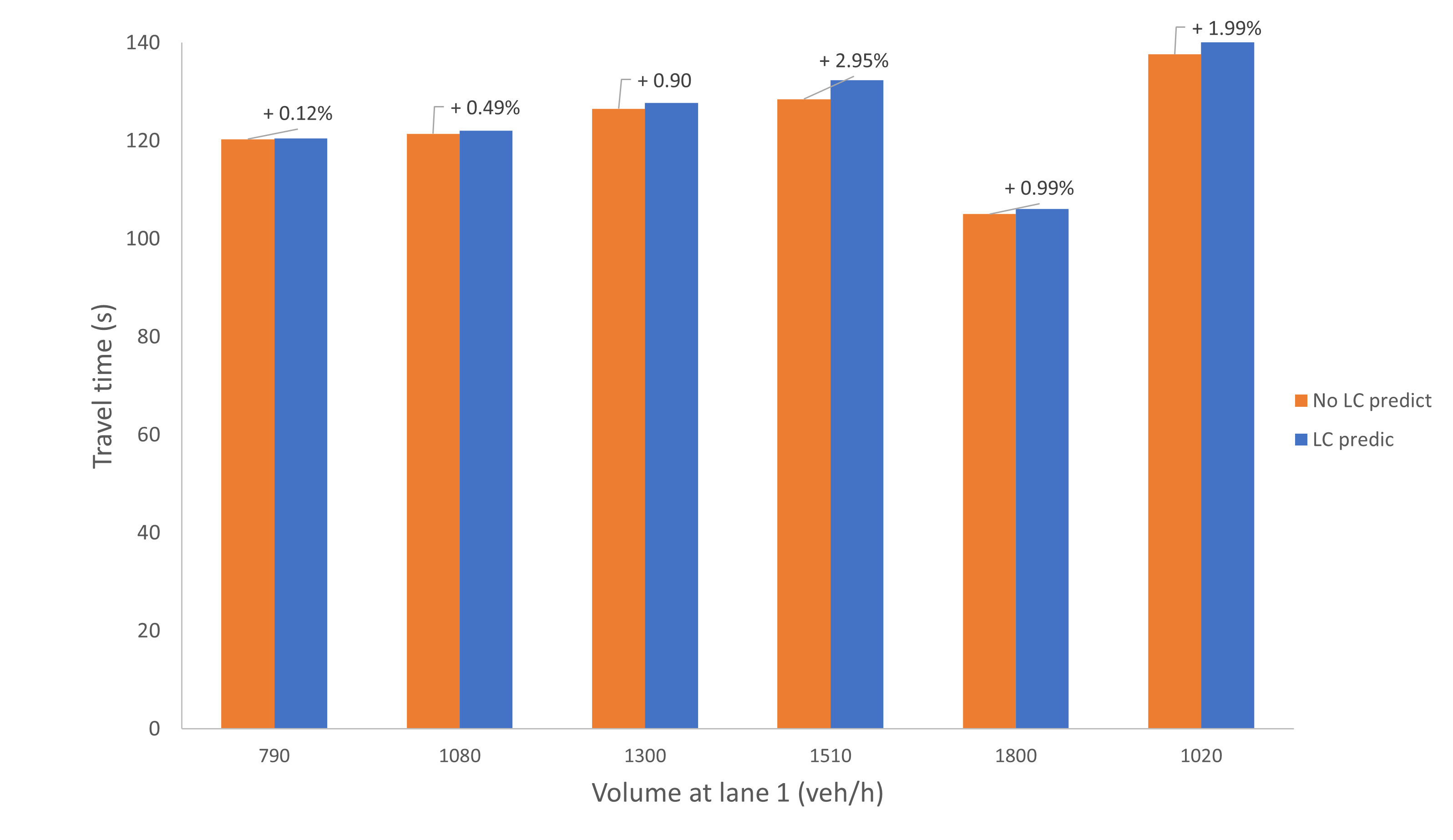}
    \caption{Travel time comparison in various volumes at 70\% MPR}
    \label{traveltflows}
\end{figure}

In the previous sections, it is presented that the modified framework results in more energy efficiency at various cases. To assess the impact of our proposed framework on travel times, we recorded the travel times for various traffic volumes. Figure \ref{traveltflows} illustrates the average travel times for the modified framework and the baseline where changes are also mentioned in percentage. Notably, the travel time increases are minimal, with the average increase rate of 1.2\% of the total travel time for the target vehicles. This indicates that our framework maintains mobility, particularly for volumes equal to or less than 1300 vehicles per hour.

\subsection{Trajectory comparison}

\begin{figure}%
    \centering
    \subfloat[\centering Trajectory  \label{traj}]{{\includegraphics[width=0.45\textwidth]{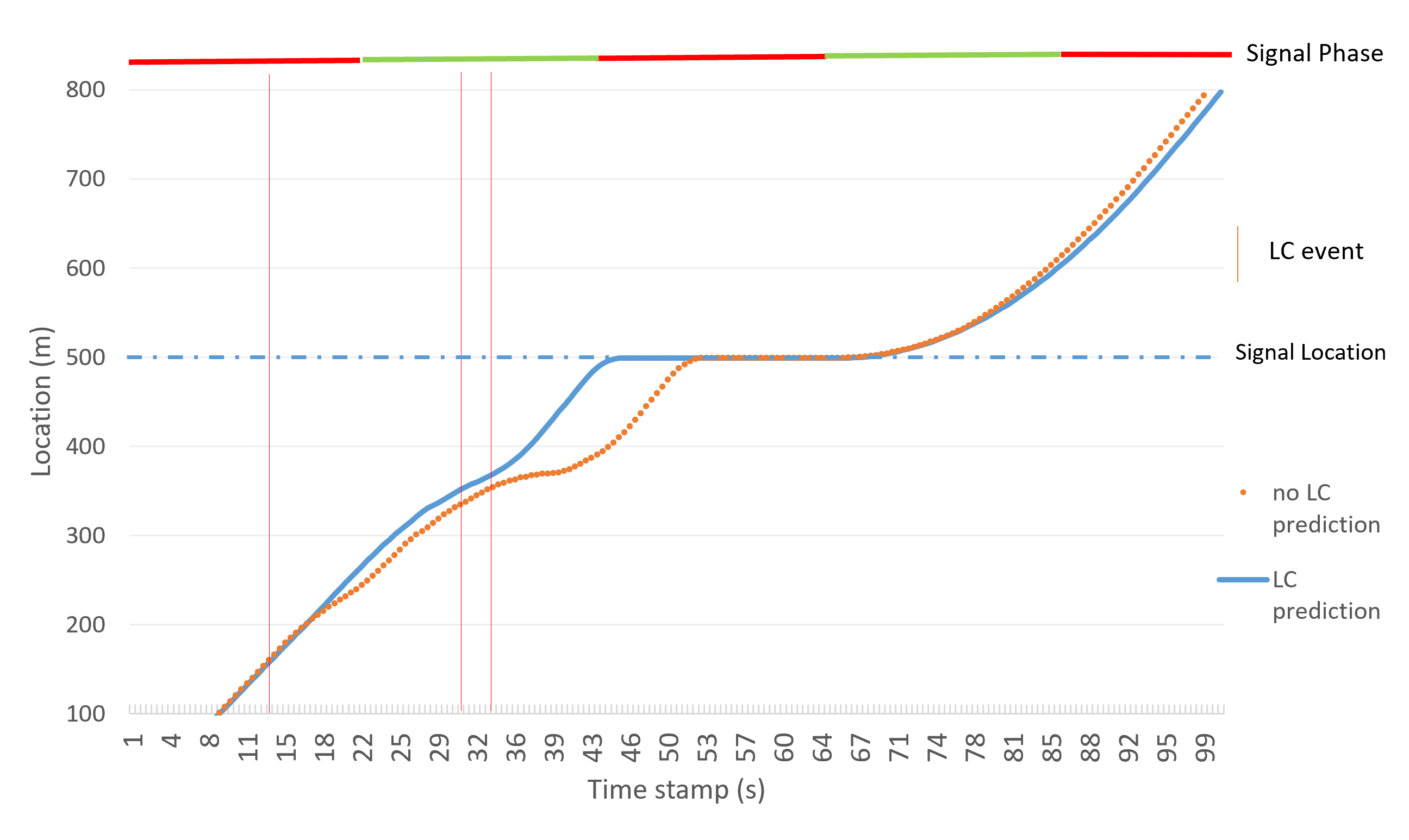} }}%
    \qquad
    \subfloat[\centering Speed plot  \label{spdplot}]{{\includegraphics[width=0.45\textwidth]{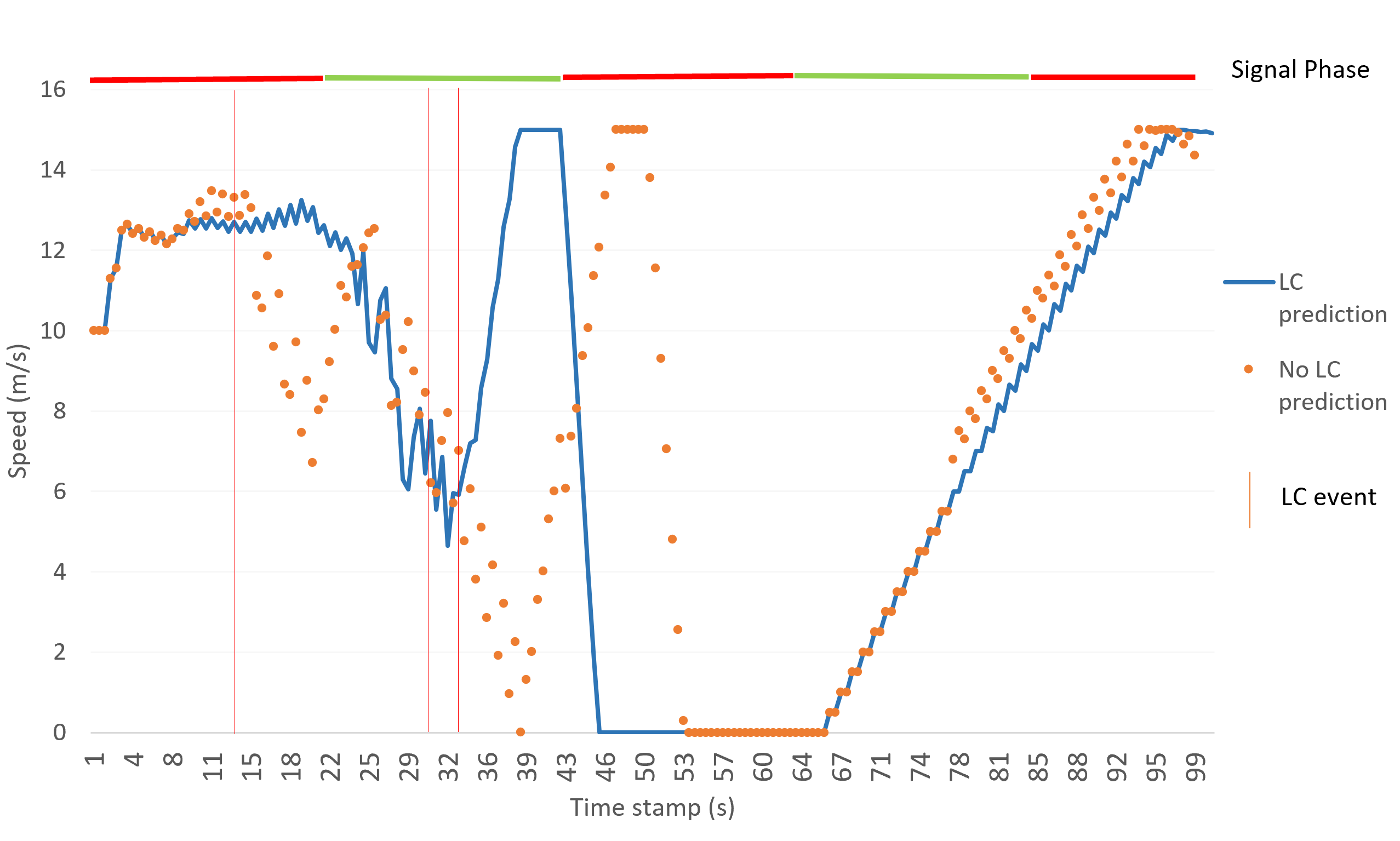} }}%
    \caption{\textcolor{blue}{Trajectory and speed comparison at 70\% MPR and volume of 1300 vehicles per hour for the target CAV}}%
    \label{trajectory}%
\end{figure}

The optimal control computes speed values for the vehicle at each time step, which are determined by the platform using Equation \eqref{eqdynamcisspd}. To assess the traveled distance over time, we plotted the trajectory and speed values recorded from both the modified framework and the baseline in Figure \ref{trajectory}. \textcolor{blue}{This figure demonstrates a single scenario with an energy benefit of 22\%, which is higher than the average value observed in the scenarios at 70\% MPR}. Figure \ref{traj} represents trajectory and Figure \ref{spdplot} represents the plot of speed as a function of time. Vertical red lines in the figure indicate the locations where LC occurred, and it is evident that the modified framework aims to smoothly reduce the speed (slope in the figure) before each lane change event. A comparison between the two trends shows that the slope changes in the non-lane change framework are larger than those in the modified framework, indicating the source of higher fuel consumption in the baseline. The vehicle controlled with LC prediction arrives at the end of the scenario 0.5 seconds earlier.

In order to verify and evaluate the speed control, speed as a function of time is also plotted for the same scenario in Figure \ref{spdplot} with the locations of the LC. It can be observed that the LC predicted scenario (orange line) avoids speed increase and even starts to reduce the speed earlier than the baseline at the points where LC happens. Speed control plans after the red light are almost the same since no LC is predicted at that time.

\textcolor{blue}{It should be noted that this is an optimistic scenario of employing LC prediction since the classic speed control (baseline) is misled when two lane changes happen between around time 29 and 32 seconds. The baseline has not predicted LC in advance and suddenly faces two new preceding vehicles. That is the main reason the target vehicle slows down to almost a full stop when there is no LC prediction (around $t=39$) in the baseline (two lane-changing vehicles were HVs, therefore the baseline control was unaware of their maneuver). Error in prediction causes higher energy consumption that our proposed framework can address. The speed control with enabled LC prediction has estimated the LC occurrence and has gradually slowed down (lower speed values starting from time 25 until time 35 help the vehicle to avoid a harsh brake when a vehicle cuts in front of the target CAV. Skipping this major slowdown is the main source of energy benefits.}

\subsection{Traffic prediction accuracy evaluation}
The presented energy benefits are attributed to the entire framework, which includes the LC prediction model working in conjunction with the modified traffic flow model. However, it is beneficial to assess the accuracy of the modified traffic model separately from the speed control. To evaluate this, it is assumed that the time and location of lane changes are predetermined for various traffic volumes. Therefore, the LC density for Equations \eqref{eqmodifyrho} and \eqref{eqmodifyv} is given for cells. 

The true values are retrieved from SUMO and both modified PW model and the standard PW model are compared to this reference. The comparison between the predicted values and the real values is conducted using the root mean square error (RMSE) metric: 

\begin{equation}
    RMSE_{\mathrm{speed}} = \sqrt{\frac{1}{N_c} \sum_{i=1}^{N_c} (V_j(i) - \hat{V_j(i)})^2 }
\end{equation}

\begin{equation}
    RMSE_{\mathrm{density}} = \sqrt{\frac{1}{N_c} \sum_{i=1}^{N_c} (\rho_j(i) - \hat{\rho_j(i)})^2 }
\end{equation}
where $N_c$ is the number of cells in the range of the target CAV, $V_j(i)$ is the actual speed of the cell $j$ coming from SUMO simulation and $\hat{V_j(i)}$ is the predicted speed of the cell $j$ from the traffic flow prediction. Similar to speed, $\rho_j(i)$ is the actual density, and $\hat{\rho_j(i)})$ is the predicted density. 

The modified traffic flow model performance is evaluated over ten random scenarios including various flow volumes. Two performance evaluation approaches were utilized: average RMSE for all cells within the connectivity range and a specific focus on cells involved in lane changes. The results presented in Table \ref{tablestateperf} indicate that the modified model outperforms the standard model, particularly in density estimation.
\begin{table}[pos=htp!]
\begin{center}
\caption{\textcolor{blue}{Prediction performance evaluation for the modified PW model in a 2 lane network, over 200 seconds of simulated time at 70\% MPR}}
\resizebox{\linewidth}{!}{%
\begin{tabular}{ |c|c|c|c|c| } 
  \hline
  Parameter & \multicolumn{1}{|p{3cm}|}{\centering All cell baseline} & \multicolumn{1}{|p{3.6cm}|}{\centering All cell w/ LC prediction ( \% improvement)} & \multicolumn{1}{|p{3.6cm}|}{\centering LC cell baseline} & \multicolumn{1}{|p{3cm}|}{\centering LC cell w/ LC prediction ( \% improvement)} \\ 
  \hline
  
  Density (veh/cell)  & \multicolumn{1}{c}{ 0.49 } \vline & \multicolumn{1}{c}{0.465 (+5.1\%)} \vline & \multicolumn{1}{c}{ 0.76} \vline & \multicolumn{1}{c}{0.61 (+19.7\%)} \vline \\ 
  \hline
  Speed (m/s)  & \multicolumn{1}{c}{ 3.89} \vline & \multicolumn{1}{c}{3.82 (+1.8\%)} \vline & \multicolumn{1}{c}{ 3.06} \vline & \multicolumn{1}{c}{2.93 (+4.2\%)} \vline \\ 
  \hline  
\end{tabular}}
\label{tablestateperf}
\end{center}
\end{table}

\section{Conclusions}
In this work, we propose a modified traffic flow model that enhances the energy efficiency of CAVs by incorporating lane-change prediction. A traffic flow model based on second-order PW model is derived by including the flow of lane changes between adjacent lanes. An analytical model is employed to update the traffic flow predictions in cases involving lane changes. This model estimates the vehicle's location and predicts lane changes that occur solely due to speed gains. The process begins with the utilization of traffic cell states, followed by the computation of the quantitative gain resulting from lane changes. If this gain surpasses a specified threshold, the model predicts a lane change within the prediction horizon and updates the modified traffic flow model to account for the impact of this lane change. Model parameters are evaluated and presented to calibrate the prediction performance. Finally, the updated traffic states serve as constraints in the optimal control framework, where the primary objective is to reduce energy consumption. The entire framework is implemented within the SUMO simulation environment to optimally control the target CAVs and report their energy consumption. The energy improvement over LC model threshold values, various CAV penetration rates, and different traffic flows are presented. Furthermore, the corresponding travel times and prediction accuracy are evaluated. The simulations indicate energy benefits of up to an average of 13\% (Figure \ref{picMPR_benefit}) at relatively high MPRs and even demonstrate energy savings at lower MPRs.

In the presented study, certain assumptions were declared while focusing on a single intersection. Examining more complex scenarios and conducting a broader study of the problem would be helpful in representing more realistic projects. \textcolor{blue}{ Furthermore, predicting additional motivations for lane changing and studying the impact of controlled CAV on the following traffic would be worthwhile for future studies}, such as considering this framework for a network with possible re-routing. Last but not least, the LC prediction platform is developed to work in cell-based traffic models. There are still opportunities for improving accuracy within this platform.

\section*{Acknowledgements}
The authors would like to thank Shi'an Wang (Assistant Professor, Electrical and Computer Engineering, The University of Texas at El Paso) for the framework explanation and support.

\bibliography{ref}

\begin{thebibliography}{10}
\expandafter\ifx\csname url\endcsname\relax
  \def\url#1{\texttt{#1}}\fi
\expandafter\ifx\csname urlprefix\endcsname\relax\def\urlprefix{URL }\fi
\expandafter\ifx\csname href\endcsname\relax
  \def\href#1#2{#2} \def\path#1{#1}\fi

\bibitem{transportationfuelshare}
IEA, \href{https://www.iea.org/reports/global-energy-review-2021}{Global energy review 2021}, Tech. rep. (2022).
\newline\urlprefix\url{https://www.iea.org/reports/global-energy-review-2021}

\bibitem{transportationCO2share}
E.~Outlook, \href{https://www.bp.com/content/dam/bp/business-sites/en/global/corporate/pdfs/energy-economics/statistical-review/bp-stats-review-2022-full-report.pdf}{Bp energy outlook. 2022. 2022 71st edition}, Tech. rep. (2022).
\newline\urlprefix\url{https://www.bp.com/content/dam/bp/business-sites/en/global/corporate/pdfs/energy-economics/statistical-review/bp-stats-review-2022-full-report.pdf}

\bibitem{ShaoandSunb}
Y.~Shao, Z.~Sun, Vehicle speed and gear position co-optimization for energy-efficient connected and autonomous vehicles, IEEE Transactions on Control Systems Technology 29~(4) (2021) 1721--1732.
\newblock \href {https://doi.org/10.1109/TCST.2020.3019808} {\path{doi:10.1109/TCST.2020.3019808}}.

\bibitem{Chalaki}
B.~Chalaki, A.~A. Malikopoulos, Optimal control of connected and automated vehicles at multiple adjacent intersections, IEEE Transactions on Control Systems Technology 30~(3) (2022) 972--984.
\newblock \href {https://doi.org/10.1109/TCST.2021.3082306} {\path{doi:10.1109/TCST.2021.3082306}}.

\bibitem{hookeroptimal}
J.~N. Hooker, Optimal driving for single-vehicle fuel economy, Transportation Research Part A: General 22~(3) (1988) 183--201.

\bibitem{marsden2001towards}
G.~Marsden, M.~McDonald, M.~Brackstone, Towards an understanding of adaptive cruise control, Transportation Research Part C: Emerging Technologies 9~(1) (2001) 33--51.

\bibitem{geiger2012team}
A.~Geiger, M.~Lauer, F.~Moosmann, B.~Ranft, H.~Rapp, C.~Stiller, J.~Ziegler, Team annieway's entry to the 2011 grand cooperative driving challenge, IEEE Transactions on Intelligent Transportation Systems 13~(3) (2012) 1008--1017.

\bibitem{asadi2010predictive}
B.~Asadi, A.~Vahidi, Predictive cruise control: Utilizing upcoming traffic signal information for improving fuel economy and reducing trip time, IEEE Transactions on Control Systems Technology 19~(3) (2010) 707--714.

\bibitem{xu2018cooperative}
B.~Xu, X.~J. Ban, Y.~Bian, W.~Li, J.~Wang, S.~E. Li, K.~Li, Cooperative method of traffic signal optimization and speed control of connected vehicles at isolated intersections, IEEE Transactions on Intelligent Transportation Systems 20~(4) (2018) 1390--1403.

\bibitem{wang2019model}
H.~Wang, P.~Peng, Y.~Huang, X.~Tang, Model predictive control-based eco-driving strategy for cav, IET Intelligent Transport Systems 13~(2) (2019) 323--329.

\bibitem{goniros2019using}
B.~Goñi-Ros, W.~J. Schakel, A.~E. Papacharalampous, M.~Wang, V.~L. Knoop, I.~Sakata, ..., S.~P. Hoogendoorn, Using advanced adaptive cruise control systems to reduce congestion at sags: An evaluation based on microscopic traffic simulation, Transportation Research Part C: Emerging Technologies 102 (2019) 411--426.

\bibitem{shladover2012impacts}
S.~E. Shladover, D.~Su, X.~Y. Lu, Impacts of cooperative adaptive cruise control on freeway traffic flow, Transportation Research Record 2324~(1) (2012) 63--70.

\bibitem{milanes2014modeling}
V.~Milanés, S.~E. Shladover, Modeling cooperative and autonomous adaptive cruise control dynamic responses using experimental data, Transportation Research Part C: Emerging Technologies 48 (2014) 285--300.

\bibitem{van2006impact}
B.~Van~Arem, C.~J. Van~Driel, R.~Visser, The impact of cooperative adaptive cruise control on traffic-flow characteristics, IEEE Transactions on intelligent transportation systems 7~(4) (2006) 429--436.

\bibitem{ngoduy2013instability}
D.~Ngoduy, Instability of cooperative adaptive cruise control traffic flow: A macroscopic approach, Communications in Nonlinear Science and Numerical Simulation 18~(10) (2013) 2838--2851.

\bibitem{delis2015macroscopic}
A.~I. Delis, I.~K. Nikolos, M.~Papageorgiou, Macroscopic traffic flow modeling with adaptive cruise control: Development and numerical solution, Computers \& Mathematics with Applications 70~(8) (2015) 1921--1947.

\bibitem{yi2006macroscopic}
J.~Yi, R.~Horowitz, Macroscopic traffic flow propagation stability for adaptive cruise controlled vehicles, Transportation Research Part C: Emerging Technologies 14~(2) (2006) 81--95.

\bibitem{daganzo1994cell}
C.~F. Daganzo, The cell transmission model: A dynamic representation of highway traffic consistent with the hydrodynamic theory, Transportation Research Part B: Methodological 28~(4) (1994) 269--287.

\bibitem{muralidharan2015computationally}
A.~Muralidharan, R.~Horowitz, Computationally efficient model predictive control of freeway networks, Transportation Research Part C: Emerging Technologies 58 (2015) 532--553.

\bibitem{carlson2010optimal}
R.~C. Carlson, I.~Papamichail, M.~Papageorgiou, A.~Messmer, Optimal motorway traffic flow control involving variable speed limits and ramp metering, Transportation science 44~(2) (2010) 238--253.

\bibitem{zhu2014accounting}
F.~Zhu, S.~V. Ukkusuri, Accounting for dynamic speed limit control in a stochastic traffic environment: A reinforcement learning approach, Transportation Research Part C: Emerging Technologies 41 (2014) 30--47.

\bibitem{csikos2017variable}
A.~Csikós, B.~Kulcsár, Variable speed limit design based on mode dependent cell transmission model, Transportation Research Part C: Emerging Technologies 85 (2017) 429--450.

\bibitem{han2017resolving}
Y.~Han, A.~Hegyi, Y.~Yuan, S.~Hoogendoorn, M.~Papageorgiou, C.~Roncoli, Resolving freeway jam waves by discrete first-order model-based predictive control of variable speed limits, Transportation Research Part C: Emerging Technologies 77 (2017) 405--420.

\bibitem{talebpour2013speed}
A.~Talebpour, H.~S. Mahmassani, S.~H. Hamdar, Speed harmonization: Evaluation of effectiveness under congested conditions, Transportation research record 2391~(1) (2013) 69--79.

\bibitem{gomes2006optimal}
G.~Gomes, R.~Horowitz, Optimal freeway ramp metering using the asymmetric cell transmission model, Transportation Research Part C: Emerging Technologies 14~(4) (2006) 244--262.

\bibitem{lo2001dynamic}
H.~K. Lo, E.~Chang, Y.~C. Chan, Dynamic network traffic control, Transportation Research Part A: Policy and Practice 35~(8) (2001) 721--744.

\bibitem{alislam2020realtime}
S.~B. Al~Islam, A.~Hajbabaie, H.~A. Aziz, A real-time network-level traffic signal control methodology with partial connected vehicle information, Transportation Research Part C: Emerging Technologies 121 (2020) 102830.

\bibitem{zhu2018modeling}
F.~Zhu, S.~V. Ukkusuri, Modeling the proactive driving behavior of connected vehicles: A cell-based simulation approach, Computer-Aided Civil and Infrastructure Engineering 33~(4) (2018) 262--281.

\bibitem{chen2020effects}
R.~Chen, T.~Zhang, M.~W. Levin, Effects of variable speed limit on energy consumption with autonomous vehicles on urban roads using modified cell-transmission model, Journal of Transportation Engineering, Part A: Systems 146~(7) (2020) 04020049.

\bibitem{levin2016multiclass}
M.~W. Levin, S.~D. Boyles, A multiclass cell transmission model for shared human and autonomous vehicle roads, Transportation Research Part C: Emerging Technologies 62 (2016) 103--116.

\bibitem{ecoapproach}
Y.~Shao, Z.~Sun, Eco-approach with traffic prediction and experimental validation for connected and autonomous vehicles, IEEE Transactions on Intelligent Transportation Systems 22~(3) (2021) 1562--1572.
\newblock \href {https://doi.org/10.1109/TITS.2020.2972198} {\path{doi:10.1109/TITS.2020.2972198}}.

\bibitem{LWRpredict}
K.-C. Chu, R.~Saigal, K.~Saitou, Real-time traffic prediction and probing strategy for lagrangian traffic data, IEEE Transactions on Intelligent Transportation Systems 20~(2) (2019) 497--506.
\newblock \href {https://doi.org/10.1109/TITS.2018.2818686} {\path{doi:10.1109/TITS.2018.2818686}}.

\bibitem{messmer1990metanet}
A.~Messmer, M.~Papageorgiou, Metanet: A macroscopic simulation program for motorway networks, Traffic engineering \& control 31~(9) (1990).

\bibitem{zhang2019energy}
F.~Zhang, X.~Hu, R.~Langari, D.~Cao, Energy management strategies of connected hevs and phevs: Recent progress and outlook, Progress in Energy and Combustion Science 73 (2019) 235--256.

\bibitem{dong2022practical}
P.~Dong, J.~Zhao, X.~Liu, J.~Wu, X.~Xu, Y.~Liu, S.~Wang, W.~Guo, Practical application of energy management strategy for hybrid electric vehicles based on intelligent and connected technologies: Development stages, challenges, and future trends, Renewable and Sustainable Energy Reviews 170 (2022) 112947.

\bibitem{jiang2017eco}
H.~Jiang, J.~Hu, S.~An, M.~Wang, B.~B. Park, Eco approaching at an isolated signalized intersection under partially connected and automated vehicles environment, Transportation Research Part C: Emerging Technologies 79 (2017) 290--307.

\bibitem{sun2020optimal}
C.~Sun, J.~Guanetti, F.~Borrelli, S.~J. Moura, Optimal eco-driving control of connected and autonomous vehicles through signalized intersections, IEEE Internet of Things Journal 7~(5) (2020) 3759--3773.

\bibitem{du2022comfortable}
Y.~Du, J.~Chen, C.~Zhao, C.~Liu, F.~Liao, C.-Y. Chan, Comfortable and energy-efficient speed control of autonomous vehicles on rough pavements using deep reinforcement learning, Transportation Research Part C: Emerging Technologies 134 (2022) 103489.

\bibitem{zhou2020stabilizing}
Y.~Zhou, S.~Ahn, M.~Wang, S.~Hoogendoorn, Stabilizing mixed vehicular platoons with connected automated vehicles: An h-infinity approach, Transportation Research Part B: Methodological 132 (2020) 152--170.

\bibitem{wei2022co}
X.~Wei, J.~Leng, C.~Sun, W.~Huo, Q.~Ren, F.~Sun, Co-optimization method of speed planning and energy management for fuel cell vehicles through signalized intersections, Journal of Power Sources 518 (2022) 230598.

\bibitem{liu2022bi}
B.~Liu, C.~Sun, B.~Wang, W.~Liang, Q.~Ren, J.~Li, F.~Sun, Bi-level convex optimization of eco-driving for connected fuel cell hybrid electric vehicles through signalized intersections, Energy 252 (2022) 123956.

\bibitem{energyandmobility}
W.~Sun, S.~Wang, Y.~Shao, Z.~Sun, M.~W. Levin, \href{https://www.sciencedirect.com/science/article/pii/S0968090X22001978}{Energy and mobility impacts of connected autonomous vehicles with co-optimization of speed and powertrain on mixed vehicle platoons}, Transportation Research Part C: Emerging Technologies 142 (2022) 103764.
\newblock \href {https://doi.org/https://doi.org/10.1016/j.trc.2022.103764} {\path{doi:https://doi.org/10.1016/j.trc.2022.103764}}.
\newline\urlprefix\url{https://www.sciencedirect.com/science/article/pii/S0968090X22001978}

\bibitem{longtermpredict}
W.~Li, X.~J. Ban, J.~Zheng, H.~X. Liu, C.~Gong, Y.~Li, Real-time movement-based traffic volume prediction at signalized intersections, Journal of Transportation Engineering, Part A: Systems 146~(8) (2020) 04020081.

\bibitem{wang100MPR}
M.~Wang, W.~Daamen, S.~P. Hoogendoorn, B.~{van Arem}, \href{https://www.sciencedirect.com/science/article/pii/S0968090X13002593}{Rolling horizon control framework for driver assistance systems. part i: Mathematical formulation and non-cooperative systems}, Transportation Research Part C: Emerging Technologies 40 (2014) 271--289.
\newblock \href {https://doi.org/https://doi.org/10.1016/j.trc.2013.11.023} {\path{doi:https://doi.org/10.1016/j.trc.2013.11.023}}.
\newline\urlprefix\url{https://www.sciencedirect.com/science/article/pii/S0968090X13002593}

\bibitem{wu100MPR}
X.~Wu, X.~He, G.~Yu, A.~Harmandayan, Y.~Wang, Energy-optimal speed control for electric vehicles on signalized arterials, IEEE Transactions on Intelligent Transportation Systems 16~(5) (2015) 2786--2796.
\newblock \href {https://doi.org/10.1109/TITS.2015.2422778} {\path{doi:10.1109/TITS.2015.2422778}}.

\bibitem{hom100MPR}
B.~HomChaudhuri, A.~Vahidi, P.~Pisu, Fast model predictive control-based fuel efficient control strategy for a group of connected vehicles in urban road conditions, IEEE Transactions on Control Systems Technology 25~(2) (2017) 760--767.
\newblock \href {https://doi.org/10.1109/TCST.2016.2572603} {\path{doi:10.1109/TCST.2016.2572603}}.

\bibitem{WADUD}
Z.~Wadud, D.~MacKenzie, P.~Leiby, \href{https://www.sciencedirect.com/science/article/pii/S0965856415002694}{Help or hindrance? the travel, energy and carbon impacts of highly automated vehicles}, Transportation Research Part A: Policy and Practice 86 (2016) 1--18.
\newblock \href {https://doi.org/https://doi.org/10.1016/j.tra.2015.12.001} {\path{doi:https://doi.org/10.1016/j.tra.2015.12.001}}.
\newline\urlprefix\url{https://www.sciencedirect.com/science/article/pii/S0965856415002694}

\bibitem{treiberp81}
M.~Treiber, A.~Kesting, Traffic Flow Dynamics, Springer-Verlag Berlin Heidelberg, 2013, pp. 81--84.

\bibitem{sun2021traffic}
W.~Sun, S.~Wang, Y.~Shao, Z.~Sun, M.~W. Levin, Traffic prediction for connected vehicles on a signalized arterial (2021) 1968--1973.

\bibitem{techreport}
M.~W. Levin, Z.~Sun, S.~Wang, W.~Sun, S.~He, B.~Suh, G.~Zhao, J.~Margolis, M.~Zamanpour, Cost/benefit analysis of fuel-efficient speed control using signal phasing and timing (spat) data: Evaluation for future connected corridor deployment, Tech. rep. (2023).

\bibitem{yang2018dynamic}
D.~Yang, S.~Zheng, C.~Wen, P.~J. Jin, B.~Ran, A dynamic lane-changing trajectory planning model for automated vehicles, Transportation Research Part C: Emerging Technologies 95 (2018) 228--247.

\bibitem{wang2019review}
Z.~Wang, X.~Shi, X.~Li, Review of lane-changing maneuvers of connected and automated vehicles: Models, algorithms, and traffic impact analyses, Journal of the Indian Institute of Science 99 (2019) 589--599.

\bibitem{du2020cooperative}
R.~Du, S.~Chen, Y.~Li, J.~Dong, P.~Y.~J. Ha, S.~Labi, A cooperative control framework for cav lane change in a mixed traffic environment, arXiv preprint arXiv:2010.05439 (2020).

\bibitem{sun2021cooperative}
K.~Sun, X.~Zhao, X.~Wu, A cooperative lane change model for connected and autonomous vehicles on two lanes highway by considering the traffic efficiency on both lanes, Transportation Research Interdisciplinary Perspectives 9 (2021) 100310.

\bibitem{tao2004modeling}
R.~H. Tao, H.~Wei, Y.~Wang, V.~P. Sisiopiku, Modeling speed disturbance absorption following state-control action-expected chains: Integrated car-following and lane-changing scenarios, in: 83rd Annual Meeting of Transportation Research, Washington, DC, 2004.

\bibitem{li2022simulation}
Q.~Li, X.~Li, Z.~Huang, J.~Halkias, G.~McHale, R.~James, Simulation of mixed traffic with cooperative lane changes, Computer-Aided Civil and Infrastructure Engineering 37~(15) (2022) 1978--1996.

\bibitem{NAGALURSUBRAVETI2021103126}
H.~H.~S. {Nagalur Subraveti}, A.~Srivastava, S.~Ahn, V.~L. Knoop, B.~{van Arem}, \href{https://www.sciencedirect.com/science/article/pii/S0968090X21001455}{On lane assignment of connected automated vehicles: strategies to improve traffic flow at diverge and weave bottlenecks}, Transportation Research Part C: Emerging Technologies 127 (2021) 103126.
\newblock \href {https://doi.org/https://doi.org/10.1016/j.trc.2021.103126} {\path{doi:https://doi.org/10.1016/j.trc.2021.103126}}.
\newline\urlprefix\url{https://www.sciencedirect.com/science/article/pii/S0968090X21001455}

\bibitem{kreidieh2022lane}
A.~R. Kreidieh, Y.~Z. Farid, K.~Oguchi, Lane assignment of connected vehicles via a hierarchical system, in: 2022 IEEE 25th International Conference on Intelligent Transportation Systems (ITSC), IEEE, 2022, pp. 2372--2378.

\bibitem{gazis1962density}
D.~C. Gazis, R.~Herman, G.~H. Weiss, Density oscillations between lanes of a multilane highway, Operations Research 10~(5) (1962) 658--667.

\bibitem{knoop2012quantifying}
V.~L. Knoop, S.~Hoogendoorn, Y.~Shiomi, C.~Buisson, Quantifying the number of lane changes in traffic: Empirical analysis, Transportation research record 2278~(1) (2012) 31--41.

\bibitem{knoop2010lane}
V.~L. Knoop, A.~Duret, C.~Buisson, B.~Van~Arem, Lane distribution of traffic near merging zones influence of variable speed limits, in: 13th International IEEE Conference on Intelligent Transportation Systems, IEEE, 2010, pp. 485--490.

\bibitem{roncoli2015traffic}
C.~Roncoli, M.~Papageorgiou, I.~Papamichail, Traffic flow optimisation in presence of vehicle automation and communication systems--part i: A first-order multi-lane model for motorway traffic, Transportation Research Part C: Emerging Technologies 57 (2015) 241--259.

\bibitem{chen2023modeling}
K.~Chen, V.~L. Knoop, P.~Liu, Z.~Li, Y.~Wang, Modeling the impact of lane-changing’s anticipation on car-following behavior, Transportation research part C: emerging technologies 150 (2023) 104110.

\bibitem{LClaval}
J.~A. Laval, C.~F. Daganzo, \href{https://www.sciencedirect.com/science/article/pii/S019126150500055X}{Lane-changing in traffic streams}, Transportation Research Part B: Methodological 40~(3) (2006) 251--264.
\newblock \href {https://doi.org/https://doi.org/10.1016/j.trb.2005.04.003} {\path{doi:https://doi.org/10.1016/j.trb.2005.04.003}}.
\newline\urlprefix\url{https://www.sciencedirect.com/science/article/pii/S019126150500055X}

\bibitem{laval2004multi}
J.~A. Laval, C.~F. Daganzo, Multi-lane hybrid traffic flow model: Quantifying the impacts of lane-changing maneuvers on traffic flow (2004).

\bibitem{LCimpact1}
H.~Yao, X.~Li, \href{https://www.sciencedirect.com/science/article/pii/S0968090X21001984}{Lane-change-aware connected automated vehicle trajectory optimization at a signalized intersection with multi-lane roads}, Transportation Research Part C: Emerging Technologies 129 (2021) 103182.
\newblock \href {https://doi.org/https://doi.org/10.1016/j.trc.2021.103182} {\path{doi:https://doi.org/10.1016/j.trc.2021.103182}}.
\newline\urlprefix\url{https://www.sciencedirect.com/science/article/pii/S0968090X21001984}

\bibitem{SuiyiLC}
S.~He, S.~Wang, Y.~Shao, Z.~Sun, M.~W. Levin, Real-time traffic prediction considering lane changing maneuvers with application to eco-driving control of electric vehicles, in: 2023 IEEE Intelligent Vehicles Symposium (IV), 2023, pp. 1--7.
\newblock \href {https://doi.org/10.1109/IV55152.2023.10186645} {\path{doi:10.1109/IV55152.2023.10186645}}.

\bibitem{V2V}
F.~Hu, \href{https://books.google.com/books?id=GXJQDwAAQBAJ}{Vehicle-to-Vehicle and Vehicle-to-Infrastructure Communications: A Technical Approach}, CRC Press, 2018.
\newline\urlprefix\url{https://books.google.com/books?id=GXJQDwAAQBAJ}

\bibitem{mainUKF}
G.~A. Terejanu, Unscented kalman filter tutorial, University at Buffalo, Buffalo (2011).

\bibitem{suhUKF}
B.~Suh, Y.~Shao, Z.~Sun, Vehicle speed prediction for connected and autonomous vehicles using communication and perception, in: 2020 American Control Conference (ACC), IEEE, 2020, pp. 448--453.

\bibitem{payne1979}
H.~J. Payne, Freflo: A macroscopic simulation model of freeway traffic, Transportation Research Record~(722) (1979).

\bibitem{treiberbook}
M.~Treiber, A.~Kesting, Traffic flow dynamics, Springer, 2013.

\bibitem{timeLC}
P.~Finnegan, P.~Green, Time to change lanes: A literature review (1990).

\bibitem{treiberbook134138}
M.~Treiber, A.~Kesting, Traffic flow dynamics, Springer, 2013, pp. 134--138.

\bibitem{diffusionreference}
M.~Mandjes, J.~Storm, A diffusion-based analysis of a multiclass road traffic network, Stochastic Systems 11~(1) (2021) 60--81.

\bibitem{diffusionreference2}
F.~van Wageningen-Kessels, B.~van't Hof, S.~Hoogendoorn, J.~van Lint, C.~Vuik, Numerical diffusion in traffic flow simulations: Accuracy analysis based on the modified equation method, in: 11th TRAIL Congress, Connecting people-Integrating expertise, TRAIL Research School, 2010, pp. 1--4.

\bibitem{LCmodel}
D.~Krajzewicz, \href{https://doi.org/10.1007/978-1-4419-6142-6_7}{Traffic Simulation with SUMO -- Simulation of Urban Mobility}, Springer New York, New York, NY, 2010, pp. 269--293.
\newblock \href {https://doi.org/10.1007/978-1-4419-6142-6_7} {\path{doi:10.1007/978-1-4419-6142-6_7}}.
\newline\urlprefix\url{https://doi.org/10.1007/978-1-4419-6142-6_7}

\bibitem{SLSQPbook}
J.-F. Bonnans, J.~C. Gilbert, C.~Lemar{\'e}chal, C.~A. Sagastiz{\'a}bal, Numerical optimization: theoretical and practical aspects, Springer Science \& Business Media, 2006.

\bibitem{parametric}
A.~G. Simpson, Parametric modelling of energy consumption in road vehicles (2005).

\bibitem{pseudo-spectral}
I.~M. Ross, M.~Karpenko, \href{https://www.sciencedirect.com/science/article/pii/S1367578812000375}{A review of pseudospectral optimal control: From theory to flight}, Annual Reviews in Control 36~(2) (2012) 182--197.
\newblock \href {https://doi.org/https://doi.org/10.1016/j.arcontrol.2012.09.002} {\path{doi:https://doi.org/10.1016/j.arcontrol.2012.09.002}}.
\newline\urlprefix\url{https://www.sciencedirect.com/science/article/pii/S1367578812000375}

\bibitem{plotkin}
S.~Plotkin, D.~Santini, A.~Vyas, J.~Anderson, M.~Wang, D.~Bharathan, J.~He, Hybrid electric vehicle technology assessment: methodology, analytical issues, and interim results., Tech. rep., Argonne National Lab., IL (US) (2002).

\bibitem{sumo-traCI}
{SUMO TraCI: Traffic Control Interface}, \url{https://sumo.dlr.de/docs/TraCI.html}, accessed: [2023].

\bibitem{sumo-traCI2}
S.~R. Santana, J.~J. Sanchez-Medina, E.~Rubio-Royo, How to simulate traffic with sumo, in: Computer Aided Systems Theory--EUROCAST 2015: 15th International Conference, Las Palmas de Gran Canaria, Spain, February 8-13, 2015, Revised Selected Papers 15, Springer, 2015, pp. 773--778.

\bibitem{wang2005real}
Y.~Wang, M.~Papageorgiou, Real-time freeway traffic state estimation based on extended kalman filter: a general approach, Transportation Research Part B: Methodological 39~(2) (2005) 141--167.

\bibitem{krauss1998microscopic}
S.~Krau{\ss}, Microscopic modeling of traffic flow: Investigation of collision free vehicle dynamics (1998).

\bibitem{karbasi2023comparison}
A.~H. Karbasi, B.~B. Mehrabani, M.~Cools, L.~Sgambi, M.~Saffarzadeh, Comparison of speed-density models in the age of connected and automated vehicles, Transportation Research Record 2677~(3) (2023) 849--865.

\bibitem{ev-database}
\href{https://ev-database.org/cheatsheet/energy-consumption-electric-car}{Energy consumption of full electric vehicles}, EV Database.
\newline\urlprefix\url{https://ev-database.org/cheatsheet/energy-consumption-electric-car}

\end{thebibliography}
\makeatletter

\end{document}